\documentclass[fleqn,usenatbib]{mnras}

\usepackage{newtxtext,newtxmath}
\usepackage[T1]{fontenc}
\usepackage{ae,aecompl}

\usepackage{graphicx}	
\usepackage{amsmath}	
\usepackage{amssymb}	

\title[GASP. X: APEX detection of molecular gas in the tails and in the disks of ram-pressure stripped galaxies]{GASP. X: APEX detection of molecular gas in the tails and in the disks of ram-pressure stripped galaxies}

\author[A. Moretti et al.]{
A. Moretti,$^{1}$\thanks{E-mail: alessia.moretti@oapd.inaf.it}
R. Paladino,$^{2}$
B. M. Poggianti,$^{1}$
M. D'Onofrio,$^{3}$
D. Bettoni,$^{1}$
\newauthor
M. Gullieuszik,$^{1}$
Y. L. Jaff\'e,$^{4}$
B. Vulcani,$^{1,5}$
G. Fasano,$^{1}$
J. Fritz$^{6}$
\newauthor
and K. Torstensson$^{7}$
\\
$^{1}$INAF-Astronomical Observatory of Padova, Vicolo dell'Osservatorio 5, 35122 Padova, Italy\\
$^{2}$INAF-Istituto di Radioastronomia and Italian ALMA Regional Centre, via P. Gobetti 101, I-40129 Bologna, Italy\\
$^{3}$Department of Physics and Astronomy, University of Padova, Vicolo dell'Osservatorio 5, 35122 Padova, Italy\\
$^{4}$Instituto de F\'isica y Astronom\'ia, Universidad de Valpara\'iso, Avda. Gran
Breta\~{n}a 1111 Valpara\'iso, Chile\\
$^{5}$School of Physics, The University of Melbourne, Swanston St \& Tin Alley Parkville, VIC 3010, Australia\\
$^{6}$Instituto de Radioastronom\'ia y Astrof\'isica, UNAM, Campus Morelia, A.P. 3-72, C.P. 58089, Mexico\\
$^{7}$European Southern Observatory (ESO), Karl-Schwarzschild-Str. 2, 85748, Garching bei Munchen, Germany
}

\date{Accepted XXX. Received YYY; in original form ZZZ}

\pubyear{2018}

\begin{document}
\label{firstpage}
\pagerange{\pageref{firstpage}--\pageref{lastpage}}
\maketitle

\begin{abstract}
Jellyfish galaxies in clusters are key tools to understand environmental processes at work in dense environments. The advent of Integral Field Spectroscopy has recently allowed to study a significant sample of stripped galaxies in the cluster environment at z$\sim 0.05$, through the GAs Stripping Phenomena in galaxies with MUSE (GASP) survey.
However, optical spectroscopy can only trace the ionized gas component through the H$_{\alpha}$ emission that can be spatially resolved on kpc scale at this redshift.
The complex interplay between the various gas phases (ionized, neutral, molecular) is however yet to be understood. 
We report here the detection of large amounts of molecular gas both in the tails and in the disks of 4 jellyfish galaxies from the GASP sample with stellar masses $\sim 3.5\times 10^{10}-3\times 10^{11} M_{\odot}$, showing strong stripping. 
The mass of molecular gas that we measure in the tails amounts to several $10^9 M_{\odot}$ and the total mass of molecular gas ranges between 15 and 100 \% of the galaxy stellar mass. 
The molecular gas content within the galaxies is compatible with the one of normal spiral galaxies, suggesting that the molecular gas in the tails has been formed in-situ.
We find a clear correlation between the ionized gas emission $\rm H\alpha$ and the amount of molecular gas.
The CO velocities measured from APEX data are not always coincident with the underlying $\rm H\alpha$ emitting knots, and the derived Star Formation Efficiencies appear to be very low.

\end{abstract}




\section{Introduction}
Galaxy evolution is strictly linked both to the environment in which galaxies reside and to their mass \citep{dressler80,Cowie+1996,Peng+2010}. 
In particular, galaxies in clusters are more efficiently quenched than their analogous in the field \citep{Tanaka+2004,Cucciati+2006,Poggianti+2006,Guglielmo+2015}. Among the various mechanisms at play in dense environments \citep{BG06}, some are influencing both the gas and the stellar populations in galaxies and are due to gravitational/tidal interactions \citep{Spitzer+Baade1951,Toomre1972,Merritt1983,Byrd+Valtonen1990,moore+1999}, while others involve only the gaseous components, resulting from the hydrodynamical interaction between the hot intra--cluster medium and the gas in the disk/halo of galaxies \citep{Larson+1980,Balogh+2000}. Among the latter, ram pressure stripping (RPS, \citealt{GG72}) seems to be favored in shaping galaxy properties in clusters \citep{Chung2009,Gavazzi+2013,Jaffe2015,Yoon+2017}.

In order to build a statistically significant sample of cluster galaxies with evidence of RPS (either in action, or already in an advances stage, leaving as signatures truncated $\rm H\alpha$ disks), we started a survey of candidate jellyfish galaxies (i.e. galaxies showing optical signatures of unilateral debris/disturbed morphology, suggestive of gas-only removal processes) in low-redshift clusters  \citep{Poggianti+2016}. From this sample we identified as targets galaxies with clear optical asymmetric morphologies suggestive of ram--pressure stripping and observed them with the MUSE spectrograph at VLT (GASP\footnote{\url{http://web.oapd.inaf.it/gasp}}, GAs Stripping Phenomena in galaxies with MUSE, \citealt{gaspI}), including also in the observed sample field and groups galaxies with similar characteristics. Our first results reveal that large amount of gas can be stripped from the galaxy main body due to the interaction with the intra-cluster medium. 
$\rm H\alpha$ knots have been found in all the GASP cluster galaxies \citep{gaspI,gaspII,gaspIII,gaspIV,gaspV}, as well as in NGC 4388 \citep{Yagi+2013} and other Virgo cluster regions \citep{Gerhard+2002,Cortese+2004} as well as in other jellyfish galaxies \citep{Fumagalli2014,Fossati2016,Consolandi+2017}, and have been associated to star formation in the galaxy tails.
However, since star formation is known to take place in regions where cold gas is present, complementary observations measuring the amount and location of molecular hydrogen are needed in order to assess if there is extraplanar molecular gas and whether this gas has been formed in situ or if it has been stripped from the galaxy body \citep{Jachym2014,Kenney+1989,Boselli+1997}.
In normal galaxies the molecular gas tends to be concentrated in the central regions, displaying a surface density decreasing with radius \citep{Young+Scoville1991,Wong+Blitz2002}, while HI disks seem to be much more extended than the optical ones \citep{Bosma1981}.
RPS would then strip mainly neutral atomic gas from the disks, producing the HI morphologies detected in Virgo and Coma \citep{Chung2009,Serra2013,Kenney2004,Kenney2014}. This gas then appears as ionized and revealed through the $\rm H\alpha$ emission \citep{Gavazzi2001,Sun2007,Yagi2007,Yagi2010,Fossati2012,Yagi2017}, and finally gets heated to the cluster X--rays temperature \citep{Sun2010}.

The cold molecular gas phase has been detected so far in only four stripped galaxies mainly through CO emission \citep{Jachym2014,Verdugo2015,Jachym2017,Dasyra2012}.
Further evidence is found in the Virgo cluster, where molecular gas in the tails of a dwarf galaxy (upper limits only, \citealt{Jachym2013}) and in the vicinity of the disk of two galaxies (\citealt{Vollmer2008}) have been detected.
High-resolution CO data in three Virgo galaxies \citep{Lee+2016} have shown that the overall distribution of cold gas seems to follow the stripped HI, but no clear sign of molecular gas stripping has been detected.

The presence of CO emission in galaxy outskirts is somewhat surprising, given its overall distribution in galaxy disks, and remains unclear whether this cold gas component could have been stripped together with the neutral one, or if, instead, it has been formed locally after the gas was stripped. In fact, the survival of dense gas clouds in the Intra--Cluster Medium (ICM) is difficult to understand, unless processes like heat conduction, turbulence and the ionizing X-rays radiation show low efficiencies. 
Finally, in order to understand how RPS proceeds, we must understand the efficiency of star formation in the tails.
This is usually derived from the comparison between the surface densities of the SFR and the molecular gas content, which shows a roughly linear relation, the so called Schmidt-Kennicutt relation \citep{Schmidt1959,Kennicutt1998}.
This relation, however, is not universal, as in extreme conditions (such as galaxy centers, galaxy outskirts, low surface brightness galaxies) it can fail \citep{Casasola+2015,Dessauges2014,Boissier2008}.

Normal disk galaxies at low redshift show average depletion times (the timescale to convert all the molecular gas reservoir in a given region into stars at the current Star Formation Rate (SFR)),  of $\tau_{dep}\sim 1-2$ Gyr \citep{Bigiel2008,Bigiel2011}, while galaxies extending towards the green valley/passive region of the SFR versus stellar mass ($M_{\star}$) plane \citep{Saintonge2017} have depletion times up to $\sim 10$ Gyr. Finally, low star formation efficiencies have been found both in the extended UV disk of M63 \citep{Bigiel2010,Dessauges2014} and in low surface brightness galaxies \citep{Boissier2008}.

As for ram--pressure stripped galaxies, the detection of molecular gas along the tails is relatively recent, and indicates a low star formation efficiency.
The first detection of an abundant molecular gas component in a tail was reported by \citet{Jachym2014} in
the galaxy ESO 137-001 in the Norma cluster, with a stellar mass ($\sim 10^9 M_{\odot}$) that is comparable with the ionized gas mass. The two gas phases together almost completely account for the missing gas mass in the galaxy disk.
In the Virgo cluster, lower quantities of molecular gas have been found in the long tail ($\sim 70$ kpc) of NGC4388 \citep{Verdugo2015}, and only upper limits of CO emission have been reported for IC 3418 \citep{Jachym2013}.
More recently, \citet{Jachym2017} found large amounts ($\sim 10^9 M_{\odot}$) of molecular gas in the very long and collimated tail ($\sim 50 kpc$) of the galaxy D100 in the Coma cluster with the IRAM 30 m telescope. Somewhat surprisingly, the mass of ionized gas is 5-10 times smaller, making the cold gas the dominant component of the tail.
Moreover, the CO velocities display a gradient along the tail, and a differential displacement with respect to the ionized gas velocities, suggesting that the densest gas clumps might be less accelerated by the ram pressure than the ionized gas.

The derived depletion times in these tails are longer than the Hubble time, meaning that most of the molecular gas will ultimately join the ICM without forming stars.
As for the origin of the CO, it is speculated that at such large distances from the galaxy the CO can not have been stripped, and its origin must be due to in situ formation from the HI stripped gas. This means that probably the hot ICM is not able to prevent gas cooling and condensation along the tail.
In particular, \citet{Jachym2017} argued that while RPS might be able to directly strip gas as dense as $\sim 50 M_{\odot} pc^{-2}$ down to  1 kpc disk radius, it might fall short in stripping denser gas present in the form of Giant Molecular Clouds (GMC). If this is true,  then the GMCs would be gradually removed by the effects of RPS.
Simulations \citep{TonnesenBryan2009,TB12} predict, in fact, that the densest clouds can not be entirely removed by the ram pressure stripping, and are therefore disrupted.
The new stars born in the tails are probably created from less dense gas that has been able to cool and subsequently condense in the tail.
Overall the complex interplay between the density/temperature conditions of the stripped gas and the ICM determine both how long is the process of mixing into the ICM and the distribution of new stars in the tails.

In order to get a clearer picture of the multiple gas phases in ram-pressure stripped galaxies, we 
obtained APEX (the Atacama Pathfinder Experiment) $^{12}\rm CO(2-1)$ data for 4 galaxies of the GASP sample, JO201, JO204, JO206, and JW100 located in the clusters A85, A957, IIZW108, and A2626, respectively.
Galaxy stellar masses range from $\sim 3.55 \times 10^{10} M_{\odot}$ to $\sim 3.0 \times 10^{11} M_{\odot}$, as derived from integrated galaxy spectra extracted from the MUSE cube and using the \citet{Chabrier2003} IMF, while cluster masses \citep{Moretti+2017} range between 0.38 and 1.58 $\times 10^{15}M_{\odot}$ (see Tab.\ref{tab:mass}). 
We refer the reader to the dedicated GASP papers \citep{gaspI,gaspII,gaspIV} for more details on the sampled galaxies.
\begin{table}
\centering
\caption{Target galaxies}
\label{tab:mass}
\begin{tabular}{lllll}
Galaxy & Cluster & Stellar mass       & Redshift & Cluster mass \\

       &         & $10^{10}M_{\odot}$ &         &  $10^{15}M_{\odot}$ \\      
\hline
JO201$^a$   & A85    & 3.55               &  0.045  &   1.58 \\ 
JO204$^b$   & A957   & 4                  &  0.042  &   0.44 \\
JO206$^c$  & IIZW108 & 9                  &  0.051  &   0.38 \\
JW100  & A2626   & 30                 &  0.061  &   0.40\\
\end{tabular}
\caption{(a) \citet{gaspII}; (b) \citet{gaspIV}; (c) \citet{gaspI}}
\end{table}

\section{APEX observations}
The observations of the 4 selected galaxies of the GASP sample have been taken with the 12 m antenna 
APEX, as part of two programs E-098.B-0657A-2016 in December 2016, and E-099.B-0063A-2017, from April to July 2017.

We observed the $^{12}\rm CO(2-1)$ transition ($\nu_{rest}=230.538$ GHz), using the Swedish Heterodyne Facility Instrument (SHFI, \citep{sfhi}), and  the eXtended Fast Fourier Transform Spectrometer (XFFTS) backend,
tuned to the CO line redshifted frequencies for each target. 

At the observed frequencies (220 GHz) the FWHM of the APEX  primary beam is 28 arcsec, corresponding to scales from $\sim$25 to $\sim$30 Kpc for the different targets.   
Multiple pointings have been observed for each target, in order to cover the main body, as well as  different regions in the tails, at different distances from the galaxy. Their locations are listed in Table 1, and shown in Figures 1 to 4.
The observations have been performed in a symmetric Wobbler switching mode, with maximum separation between the ON and OFF-beam of 150 arcsec. 

Observations have been taken in different sessions with different weather conditions, with Precipitable Water Vapour (PWV) ranging from 0.7 to 3 mm.

\subsection{Data reduction}
The spectra calibrated by the APEX on-line calibration pipeline (in antenna temperature scale; T$_{\rm A}^{*}$), have been 
further reduced using the standard procedure with the program CLASS from the GILDAS software package\footnote{http://www.iram.fr/IRAMFR/GILDAS}.
Bad scans were flagged and the spectra have been aligned in the same velocity frame, and 
smoothed to a common spectral resolution of 80 km/s.
Some of the pointings show emissions strong enough to be detected with S/N>3  
also at higher spectral resolution. 
First-order baselines, defined in a line-free band about 2000 km/s wide, have been subtracted 
to the spectra.
The antenna temperatures have been converted to main-beam brightness temperatures 
(T$_{\rm mb}$=T$_{\rm A}^{*}$/$\eta_{\rm mb}$), using $\eta_{\rm mb}$=0.75.

The typical levels of rms noise obtained in channels 80 km/s wide range from 0.5 mK to 1.2 mK.

The list of the observed positions with the on-source observing times 
and the observed frequency is provided in Tab. \ref{tab:obs}.
The observed frequency corresponds to the galaxy central redshift, as derived from the MUSE datacubes.
\begin{table}
\centering
\caption{Observations log }
\label{tab:obs}
\begin{tabular}{llllll}
Gx & Pos. & RA & DEC & T$_{ON}$ & Frequency \\
\smallskip
& & & & min & MHz \\
\hline
JO201 & A & 00:41:30.32 & -09:15:45.8& 30   & 220680.2\\ 
JO201 & B & 00:41:31.07 & -09:15:23.5& 175  & 220680.2 \\ 
JO201 & C & 00:41:32.80 & -09:15:20.2& 30   & 220680.2 \\ 
JO201 & D & 00:41:32.00 & -09:15:43.8& 62   & 220680.2 \\ 
JO201 & E & 00:41:33.00 & -09:16:03.6& 32   & 220680.2 \\ 
JO201 & F & 00:41:31.31 & -09:16:05.4& 66   & 220680.2 \\ 
JO204 & A & 10:13:46.34 & -00:54:53.5& 26   & 221152.2 \\ 
JO204 & B & 10:13:46.17 & -00:54:39.4& 68   & 221152.2 \\ 
JO204 & C & 10:13:48.10 & -00:54:56.5& 145  & 221152.2 \\ 
JO206 & A & 21:13:47.6  & 02:28:28.2 & 9    & 219288.5 \\ 
JO206 & B & 21:13:45.7  & 02:28:44.2 & 27   & 219288.5 \\ 
JO206 & C & 21:13:43.8  & 02:28:26.2 & 93   & 219288.5 \\ 
JO206 & D & 21:13:42.80 & 02:28:49.2 & 139  & 219288.5 \\ 
JW100 & A & 23:36:24.55 & 21:09:05.7 & 12   & 217258.7 \\  
JW100 & B & 23:36:23.73 & 21:08:44.6 & 71   & 217258.7 \\ 
\end{tabular}
\end{table}

The peak temperature (main-beam brightness temperature, $\rm T^*_{mb}$) as well as the width and the position of the CO line have been derived initially with a single Gaussian fit, and with a multiple gaussian component where needed. CO fluxes have been calculated assuming a conversion factor of S$_\nu$/T$_{mb}$ = 39 Jy beam$^{-1}$ K$^{-1}$ for the APEX telescope.
The results of the fitting procedure are given in Tab.\ref{tab:res}.

\section{MUSE data}\label{sec:muse}
Within the GASP project \citep{gaspI}, we observed so far $\sim 100$ galaxies showing more or less pronounced signatures of gas stripping phenomena in action. 
For each galaxy we have been able to derive gas and stellar properties from the single spaxels (smoothed on a $0.7-1.3 \rm kpc$ scale, depending on the galaxy redshift), as well as from the galaxy integrated spectra.
In the following Fig.\ref{fig:JO201_1G} to \ref{fig:JW100_1G} we show in the two upper panels the white light image of each target galaxy (top left panels) and the stellar kinematics (top right panels) derived from the MUSE datacubes. 
The stellar kinematics has been measured in Voronoi binned regions with a signal--to--noise of 10 \citep{CC2003} using the pPXF software \citep{CE2004}, as described in \citet{gaspI}.
Moreover, we used the $\rm H\alpha$ images extracted from the MUSE datacubes to locate the star forming knots both in the galaxy disk and along the tails.
The knots positions have been identified as local minima on the $\rm H\alpha$ image filtered using a laplacian $+$ median kernel and their size have been estimated using a recursive outward analysis of three circular shells, as better described in \citet{gaspI}.
For each star forming knot we then extracted the integrated spectrum, and measured both velocities (and velocity dispersions) and emission line fluxes, that have been subsequently used to: a) characterize the ionization mechanism responsible for the $\rm H\alpha$ emission; b) measure the gas metallicity and c) derive the ionized gas mass (as better described in Sec.\ref{sec:sfe}).
For the purpose of this paper we used only the knots characterized by a stellar ionization mechanism (i.e. star forming or composite regions), as characterized by MUSE emission-only line ratios (\citet{gaspI}.

\section{Results}\label{sec:results}

Figs. \ref{fig:JO201_1G}, \ref{fig:JO204_1G}, \ref{fig:JO206_1G} and \ref{fig:JW100_1G} show for each galaxy (a) the white light image of the galaxy (top left), (b) the stellar kinematics (top right), (c) the star forming knots (lower left) derived from the MUSE datacubes (see Sec. \ref{sec:muse}), and (d) the CO(2-1) spectra ($T_{mb}^*$) from the APEX data.
The observed APEX pointings are overplotted on all the maps: the red circle marks the pointing superimposed to the galaxy center, while the blue ones are positioned along the stripped tails.
The grey region in the lower left panels shows the galaxy emission traced by the $\rm H\alpha$ line for spaxels with a $S/N > 5$, while
the colored circles correspond to the star forming knots described in Sec. \ref{sec:muse}.
Their color corresponds to the velocity relative to the galaxy center (and to the zero velocity of APEX observations). The red boxes in the lower right panels indicate the ionized gas velocity of the star forming regions, with an height proportional to the $\rm H\alpha$ flux, normalized using different scales for the purpose of visualization, and a width corresponding to the $\rm H\alpha$ velocity dispersion.
The APEX pointings outside of the galaxy disks extend between $\sim$ 20 kpc and $\sim$ 67 kpc (in JO206) and are superimposed to regions where at least one $\rm H\alpha$ emitting region is present.

\subsection{JO201}\label{sec:JO201}
JO201 is one of the most spectacular jellyfish galaxies revealed by the GASP survey: it is moving at high velocity in the A85 cluster mostly along the line of sight, so that in the outskirts of the disk (starting from $\sim$ 6 kpc from its center) the ionized gas is increasingly redshifted. This indicates the presence of gas trailing behind the stellar component \citep{gaspII}.
Our procedure revealed the presence of 148 $\rm H\alpha$ emitting knots, 122 of them being classified as star forming. Their median ionized gas mass is $1 \times 10^6 M_{\odot}$ and in total they account for $9.57 \times 10^7$ solar masses of ionized gas.

\begin{figure*}
\centering
\begin{minipage}{0.45\textwidth}
\includegraphics[width=1.0\textwidth]{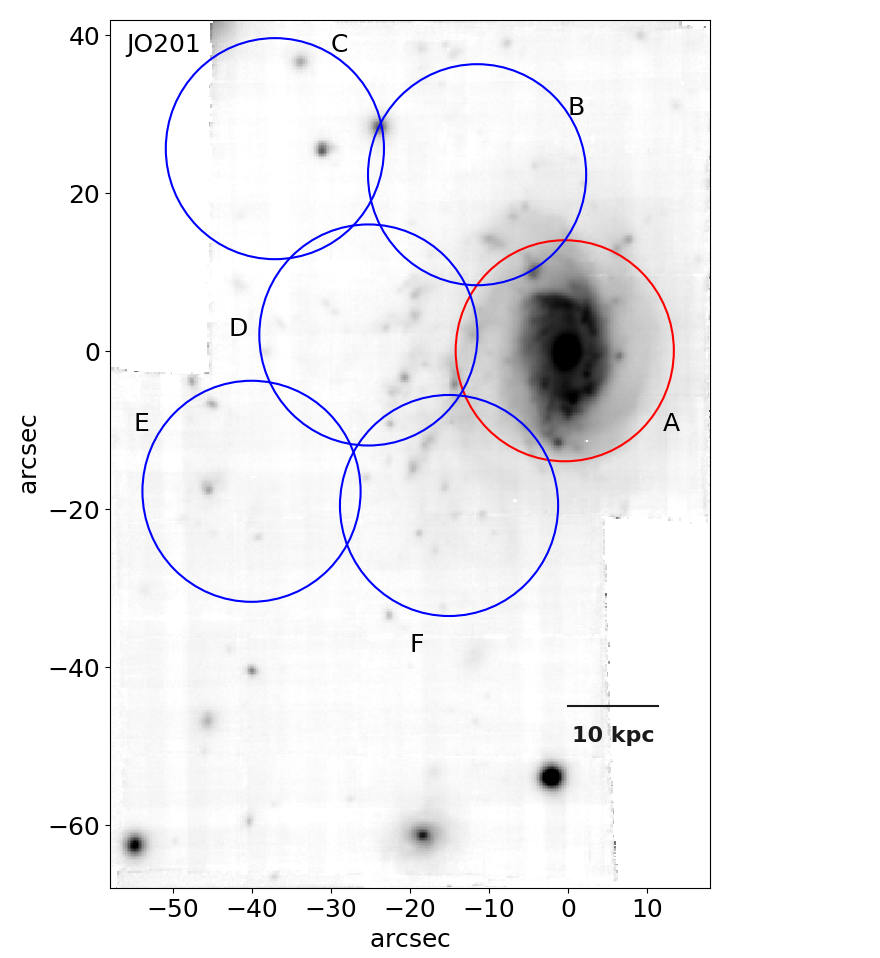}
\end{minipage}
\begin{minipage}{0.45\textwidth}
\includegraphics[width=1.0\textwidth]{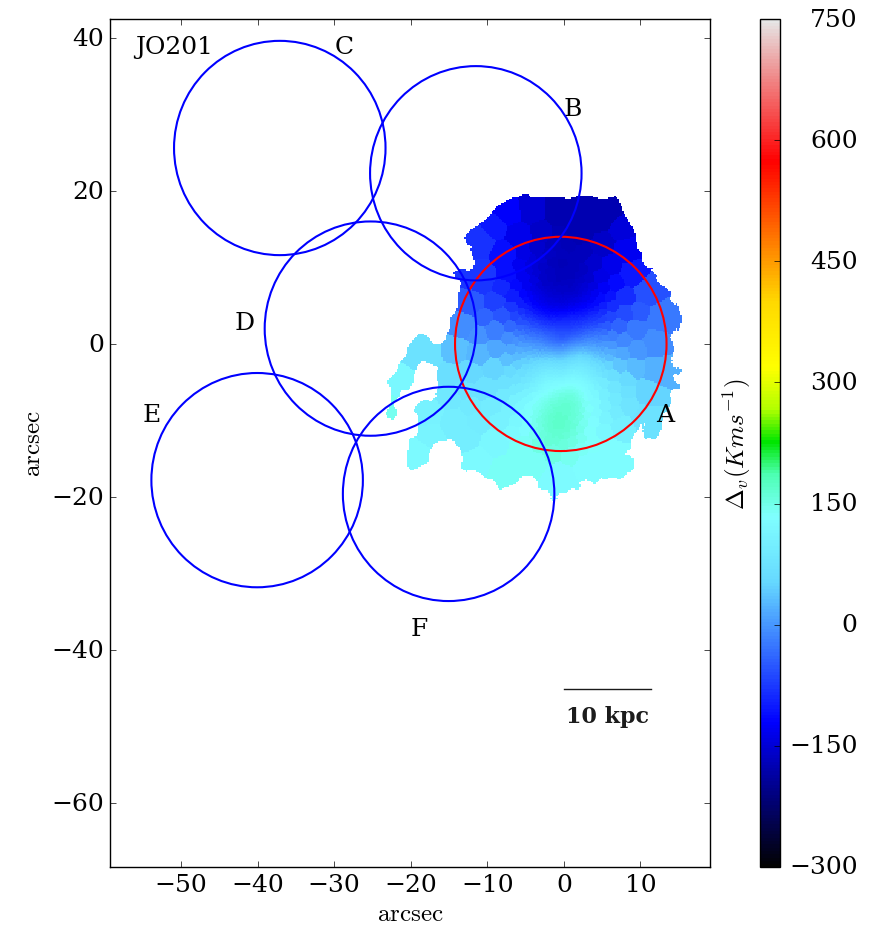}
\end{minipage}
\begin{minipage}{0.45\textwidth}
\includegraphics[width=1.0\textwidth]{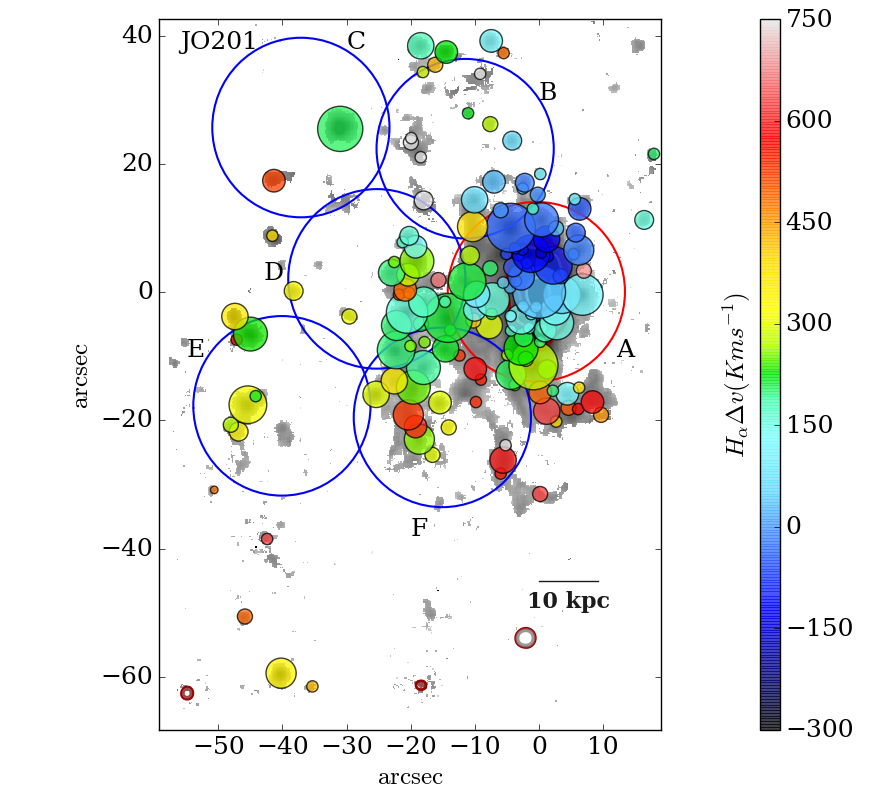}
\end{minipage}
\begin{minipage}{0.45\textwidth}
\includegraphics[width=0.45\textwidth]{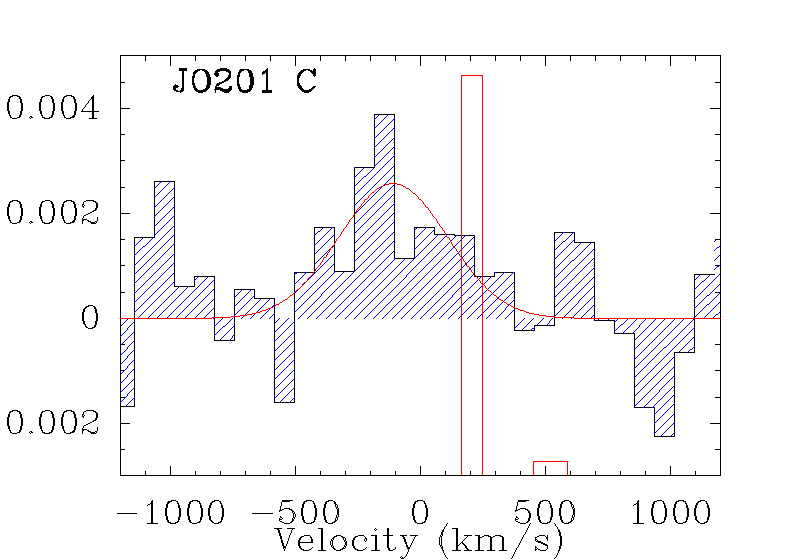}
\includegraphics[width=0.45\textwidth]{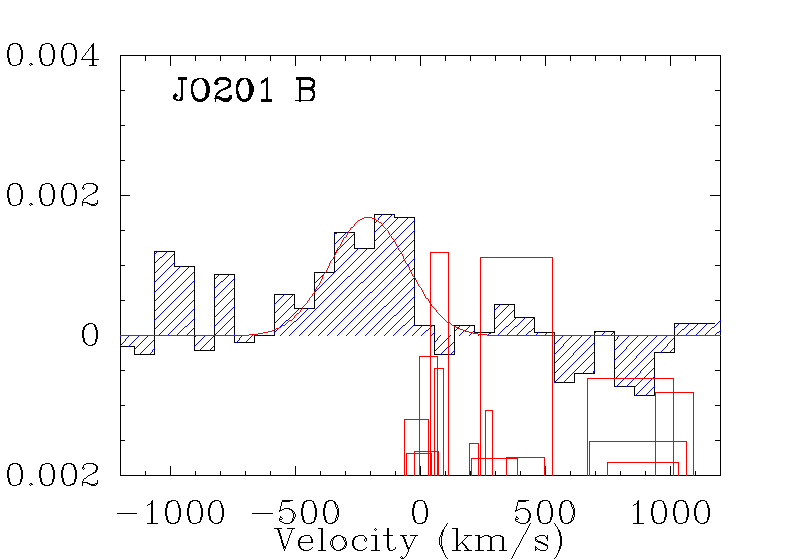}
\includegraphics[width=0.45\textwidth]{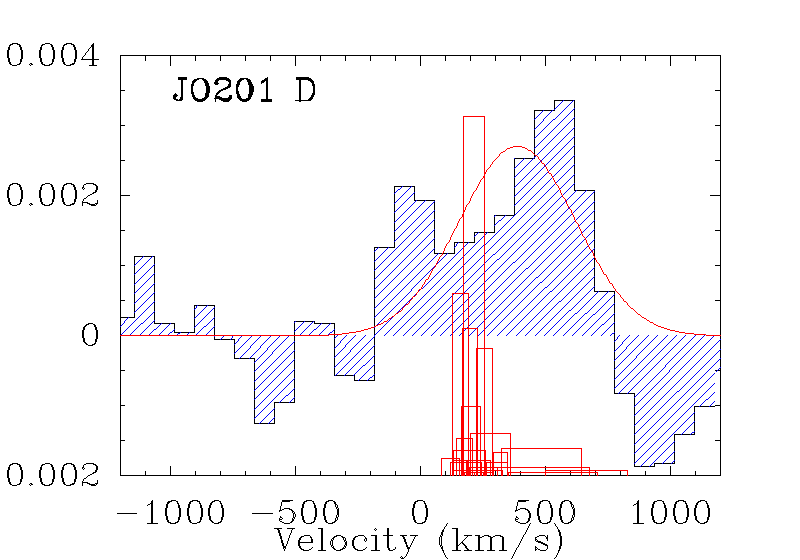}
\includegraphics[width=0.45\textwidth]{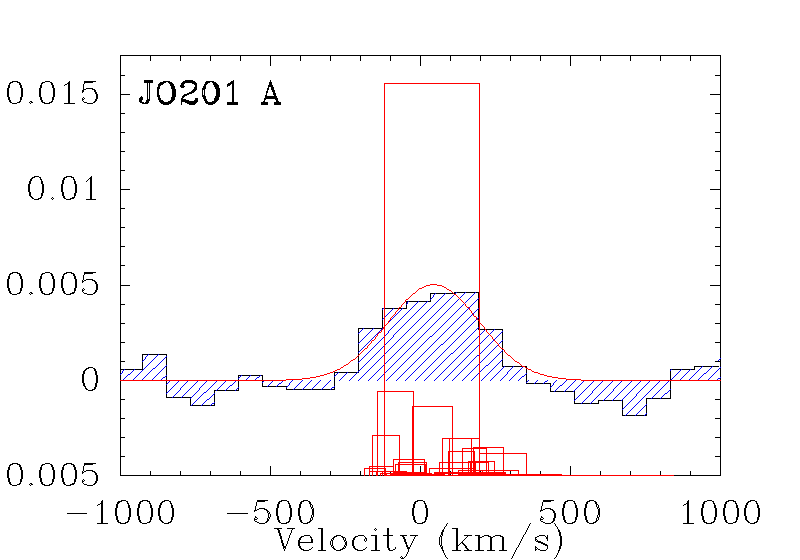}
\includegraphics[width=0.45\textwidth]{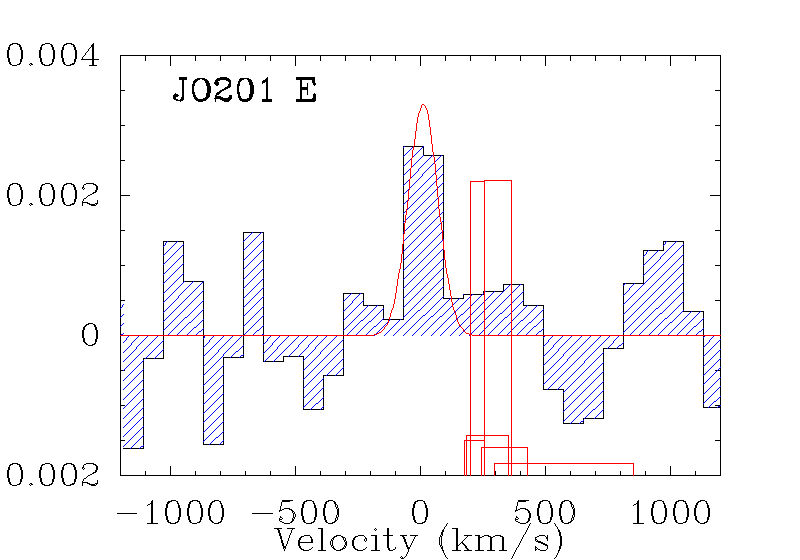}
\includegraphics[width=0.45\textwidth]{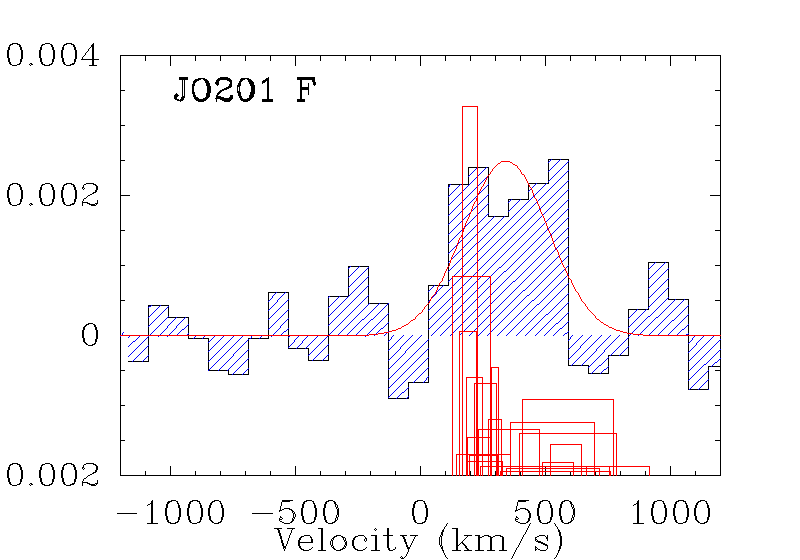}
\end{minipage}
\caption{{\bf JO201} The upper left panel shows the white light image of the galaxy extracted from the MUSE datacube. The upper right panel shows the stellar kinematics derived from MUSE spectra in Voronoi binned regions with $S/N=10$. The lower left panel shows the galaxy emission determined from the continuum under the $\rm H\alpha$ emission from the MUSE data (in grey) and the star forming knots color coded according to their velocity with respect to the galaxy center. Red (superimposed to the main galaxy body) and blue circles are the observed APEX pointings. The lower right panels show the CO($2-1$) spectra ($T^*_{mb}$) in the 6 positions. In each spectrum, the underlying red boxes corresponds to the $\rm H\alpha$ emissions of the star forming knots within the APEX beam (see text).\label{fig:JO201_1G}}
\end{figure*}

Fig.\ref{fig:JO201_1G} shows the CO detections in the six pointings of JO201.
The pointing A encompasses the galaxy center, and its CO velocity  distribution is coincident with the one of the $\rm H\alpha$ emission from the SF knots.

The velocities of the molecular gas and the ionized gas are instead only partially coincident in the pointings D and F, 
though in the D pointing the CO peak is measured with a quite low S/N (2.9, see Tab. \ref{tab:res}). The other external pointings (B, C and E) reveal the presence of molecular gas at a velocity that is different from
that of the ionized gas. In particular we were able to identify in the C and E pointings only a few HII regions/complexes, and they have a positive velocity with respect to the galaxy center, while the peak of the CO is negative (-107 km/s, pointing C) or consistent with a zero velocity (12 km/s, pointing E). Moreover, in the C pointing, where also the ionized gas is only marginally detected, the S/N of the CO detection is low (2.1). 

Given the discrete distribution of the $\rm H\alpha$ emission within the APEX beams, possibly coincident with CO concentrations,
for all the pointings except the central one we also performed a double Gaussian fit, fixing the peak velocity to the value of the corresponding $\rm H\alpha$ emission when a free fit did not recover the line. 
The results are also given in Tab. \ref{tab:res}. The asterisk marks the parameter fixed for the double fitting procedure, while the corresponding Figures are shown in the Appendix.
By imposing a double Gaussian fit the significance of the secondary peak in the D pointing increases (S/N=3.4), while it remains low in the C pointing.

The behavior of cold gas in JO201 is rather complex. The cold gas found in the main body of the galaxy as well as in the pointings closest to the center (D and F) seems to be compatible, in term of its velocity, with the ionized gas traced by the $\rm H\alpha$ emission, while more external pointings (B, C and E) show prominent cold gas emission at a different velocity with respect to the warm gas component. 

More in detail, pointing D shows a double peak, with one peak compatible with a null velocity with respect to the galaxy center, and a secondary peak (prominent in CO, less luminous in $\rm H\alpha$). A similar behavior is present also in the F pointing, that covers a region with both $\rm H\alpha$ emission at a velocity similar to the stellar velocity of the closest region on the disk, and a more redshifted $\rm H\alpha$ component that is clearly trailing behind. This seems to indicate that within the pointings superimposed on the main galaxy disk part of the molecular gas is still linked to the stellar/gaseous disk rotation, while part of it follows the stripped gas where new stars are born (traced by part of the HII knots).

In all the three more external pointings (B, C and E) the $\rm H\alpha$ emission is redshifted with respect to the center of the stellar disk, and is only marginally associated with CO peaks. Most of the CO has a velocity that is compatible with the stellar component at the closest position on the galaxy disk, as if it was less easily stripped along the line of sight. 
If this is true, then in evaluating the Star Formation Efficiency within the APEX beams, we would have to consider only the CO emitting gas associated with the $\rm H\alpha$ emission (see Sec.\ref{sec:sfe}).

\subsection{JO204}\label{sec:JO204}
\begin{figure*}
\centering
\begin{minipage}{0.45\textwidth}
\includegraphics[width=1.0\textwidth]{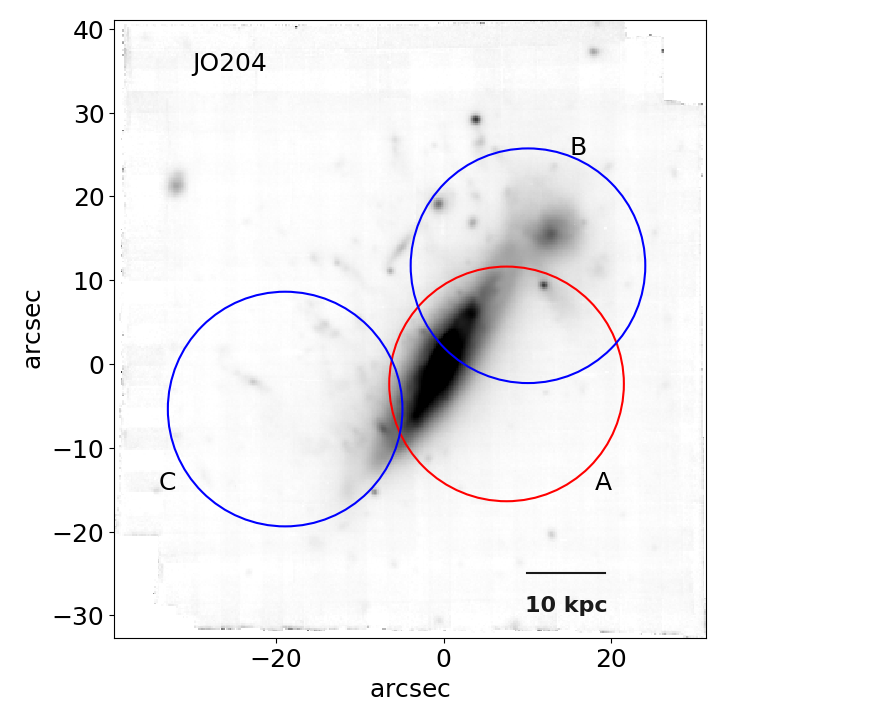}
\end{minipage}
\begin{minipage}{0.45\textwidth}
\includegraphics[width=1.0\textwidth]{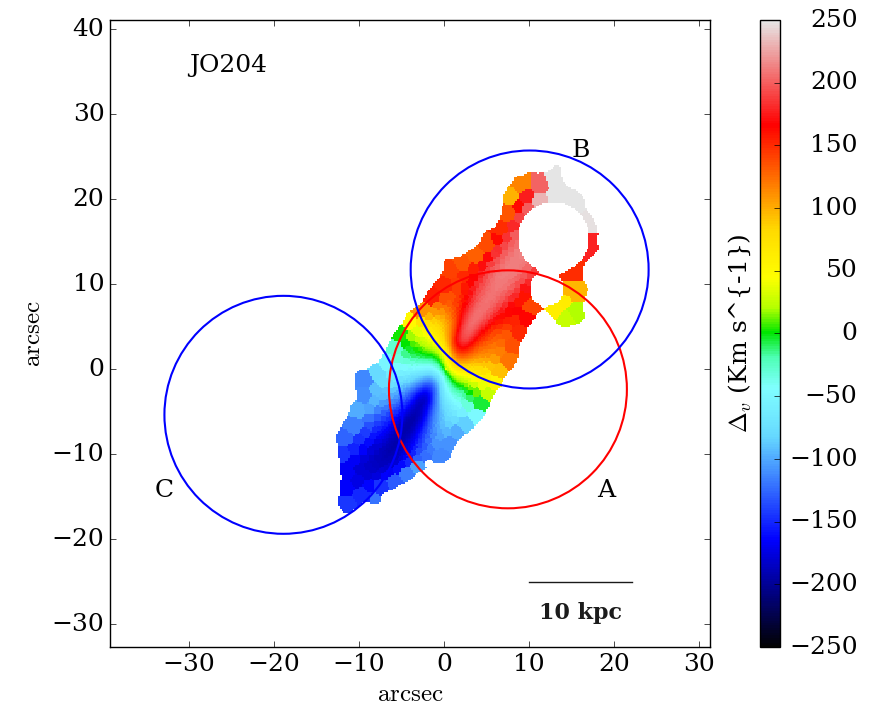}
\end{minipage}
\begin{minipage}{0.45\textwidth}
\includegraphics[width=1.0\textwidth]{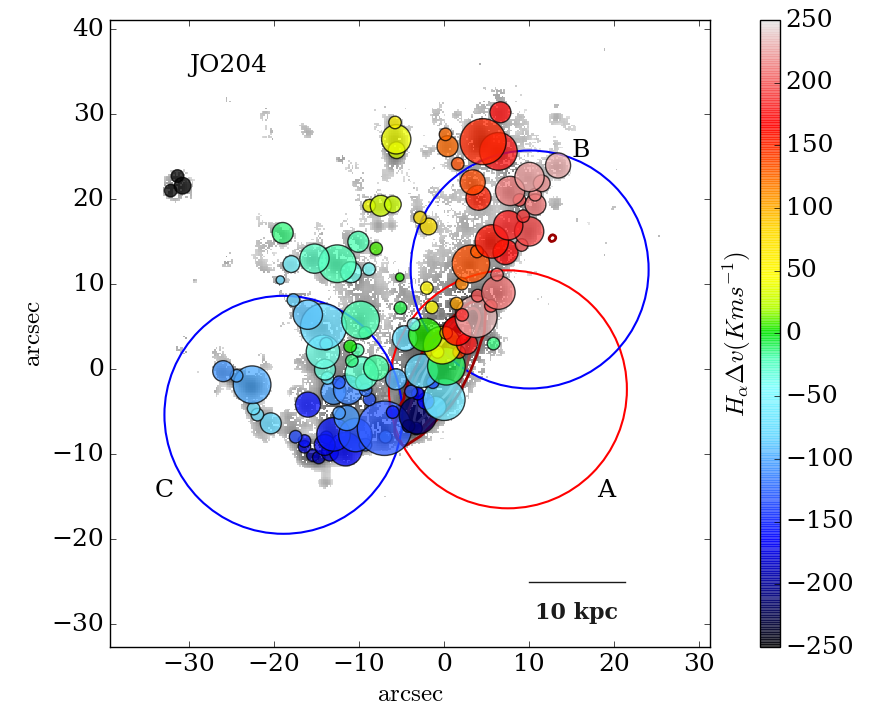}
\end{minipage}
\begin{minipage}{0.45\textwidth}
\includegraphics[width=0.44\textwidth]{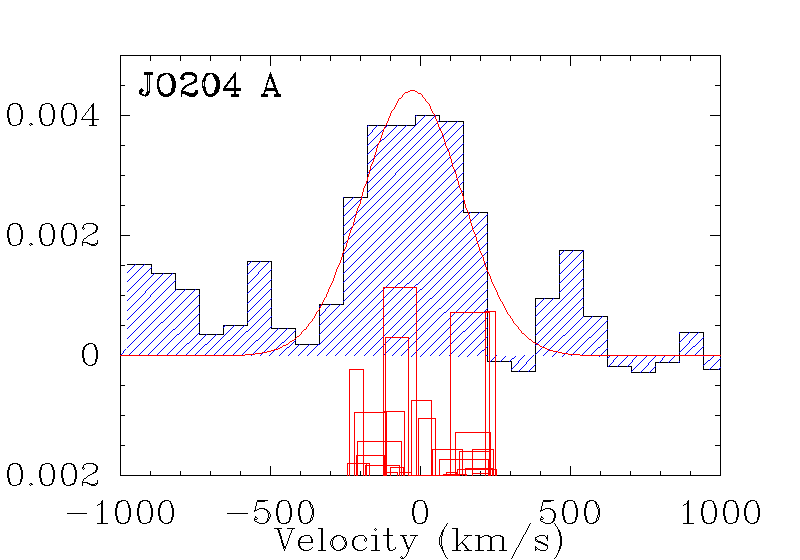}
\makebox[0.44\textwidth][c]{}
\includegraphics[width=0.44\textwidth,clip]{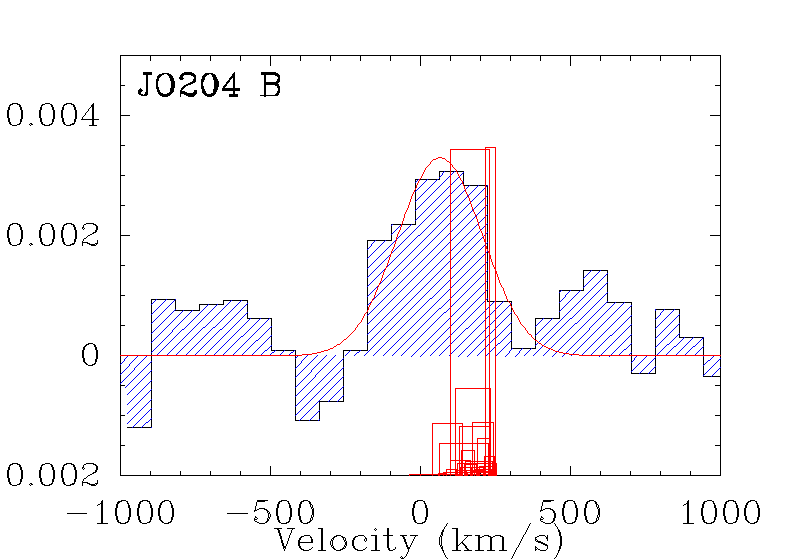}
\includegraphics[width=0.44\textwidth,clip]{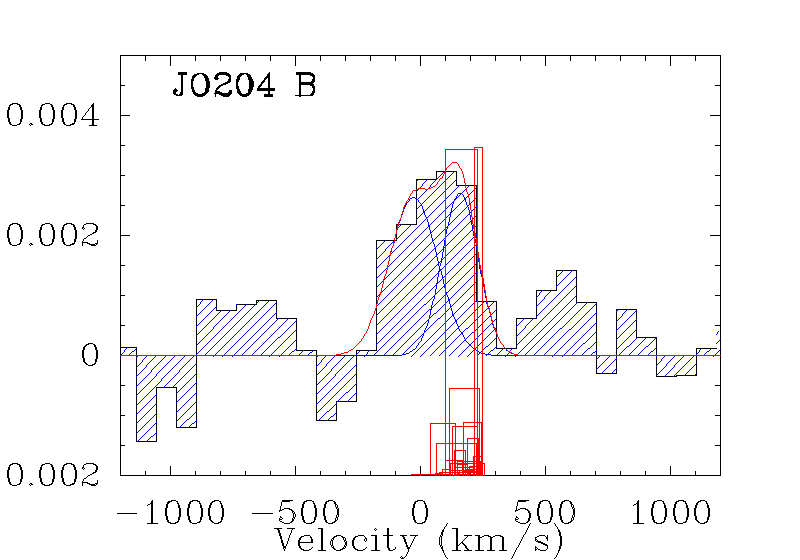}
\includegraphics[width=0.44\textwidth,clip]{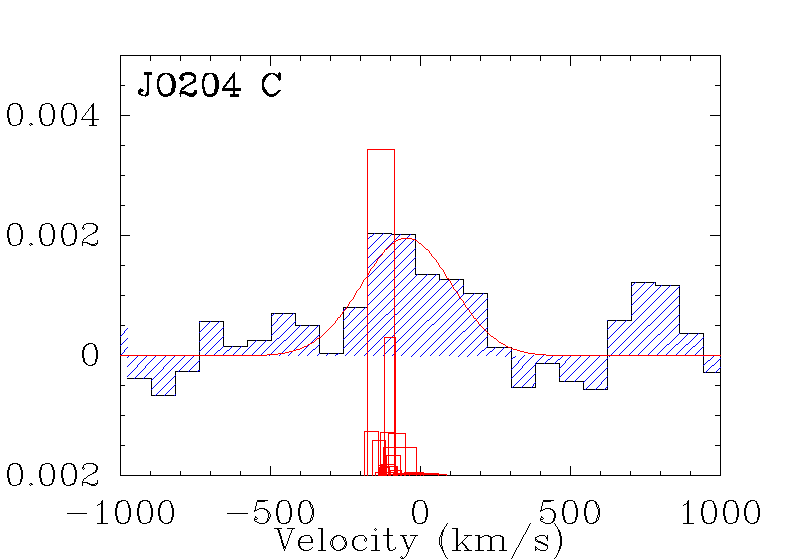}
\includegraphics[width=0.44\textwidth,clip]{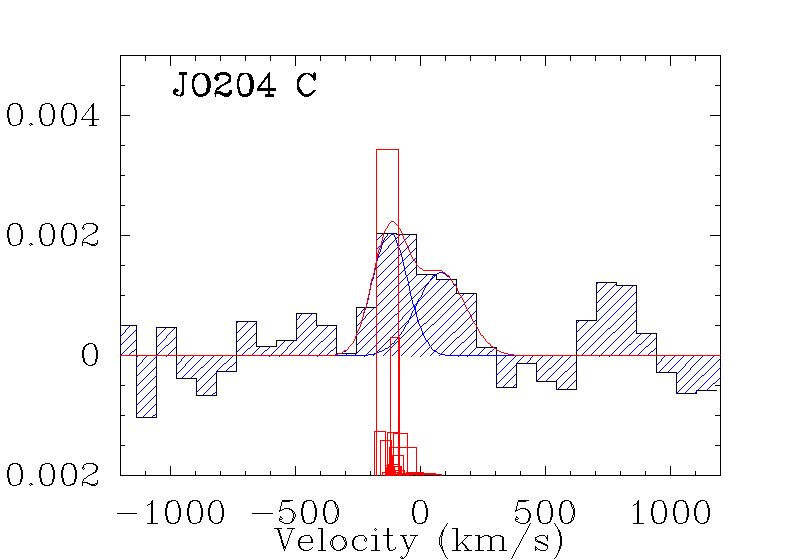}
\end{minipage}
\caption{{\bf JO204} The upper left panel shows the white light image of the galaxy extracted from the MUSE datacube. The upper right panel shows the stellar kinematics derived from MUSE spectra in Voronoi binned regions with $S/N=10$. The lower left panel shows the galaxy emission determined from the continuum under the $\rm H\alpha$ emission from the MUSE data (in grey) and the star forming knots color coded according to their velocity with respect to the galaxy center. Red (superimposed to the main galaxy body) and blue circles are the observed APEX pointings. The lower right panels show the CO($2-1$) spectra ($T^*_{mb}$) in the 3 positions. The two right panels show the double Gaussian fit. In each spectrum, the underlying red boxes corresponds to the $\rm H\alpha$ emissions of the star forming regions within the APEX beam (see text).\label{fig:JO204_1G}}
\end{figure*}

JO204 is probably at its first infall in the A957 cluster \citep{gaspIX} and is mostly being stripped on the plane of the sky, as demonstrated by the coherent rotation of the stripped gas with respect to the undisturbed stellar component (shown in the upper right panel of Fig. \ref{fig:JO204_1G}).
It possesses 92 star forming knots, according to the procedure described in Sec.\ref{sec:muse} (see also \citealt{gaspIV}), but only 51 of them have SII lines measurements within the range where the \citet{Proxauf+2014} calibration can be applied. The average ionized gas mass in these knots turns out to be $5.3 \times 10^4 M_{\odot}$, and the total ionized gas mass within these star forming knots is $1.8 \times 10^7 M_{\odot}$ (one order of magnitude smaller than the JO206 galaxy, see Sec.\ref{sec:JO206}).

As for the CO emission, shown in Fig.\ref{fig:JO204_1G}, lower right panels, it is again consistent with the overall distribution of the ionized gas in the central pointing (containing an AGN also in this case), while in the two external slightly off-disk regions the CO line is larger than the $\rm H\alpha$ emission and a double Gaussian fit (shown in the two lower right panels) reveals that the CO is again showing probably a double component: one that is coincident in velocity with the $\rm H\alpha$ emission, and another one moving at a lower absolute velocity value.

In this case the galaxy is moving in the plane of the sky and the net effect of the gas stripping on the velocity of the molecular gas is less easily seen. What we see is, in fact, the combination of the ram pressure stripping and the galaxy rotation. 

\subsection{JO206}\label{sec:JO206}
\begin{figure*}
\centering
\begin{minipage}{0.45\textwidth}
\includegraphics[width=1.0\textwidth]{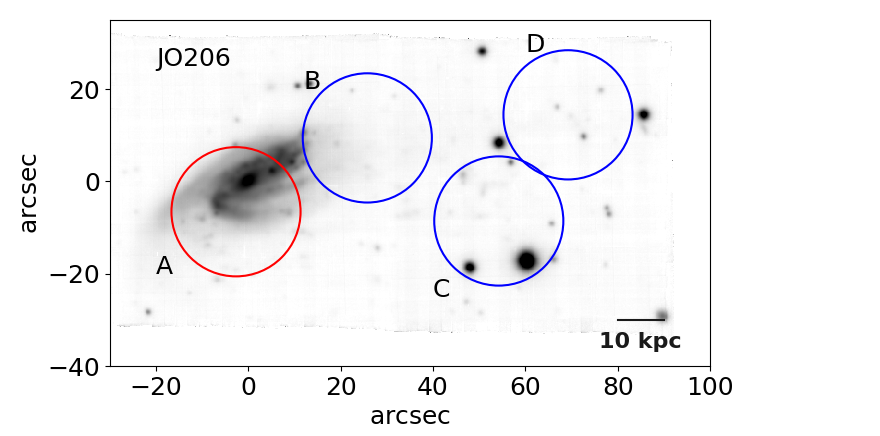}
\end{minipage}
\begin{minipage}{0.45\textwidth}
\includegraphics[width=1.0\textwidth]{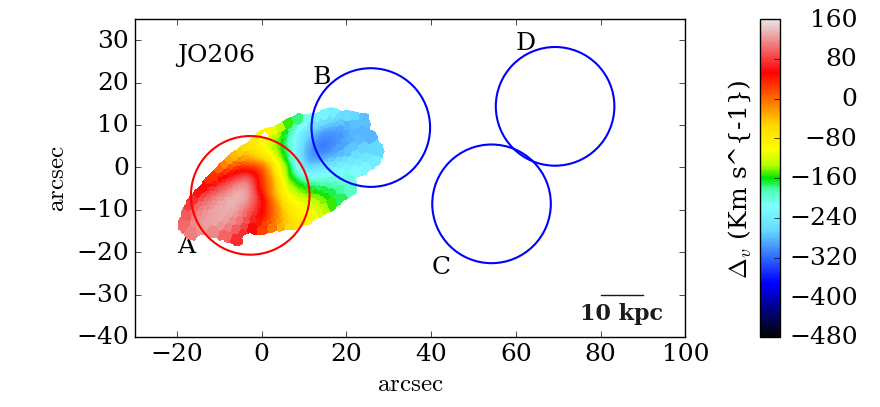}
\end{minipage}
\begin{minipage}{0.45\textwidth}
\includegraphics[width=1.0\textwidth]{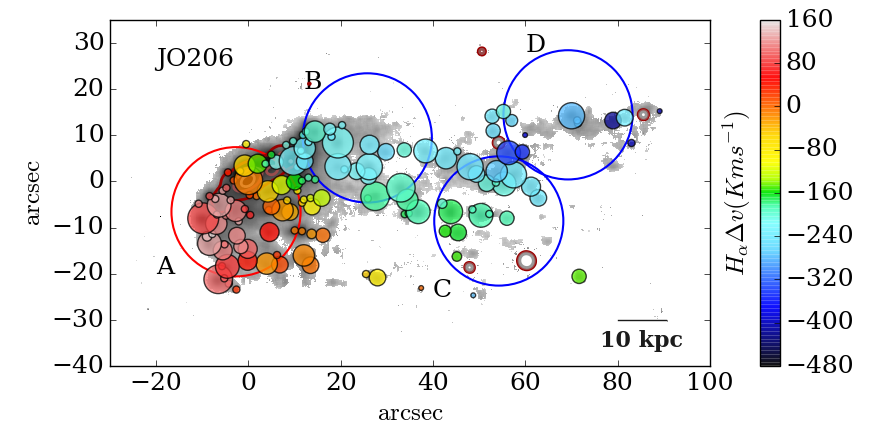}
\end{minipage}
\begin{minipage}{0.45\textwidth}
\includegraphics[width=0.45\textwidth]{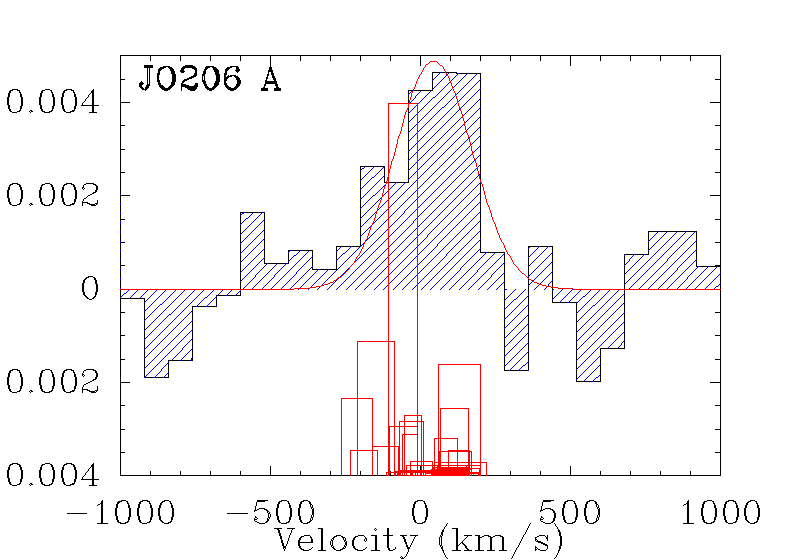}
\includegraphics[width=0.45\textwidth]{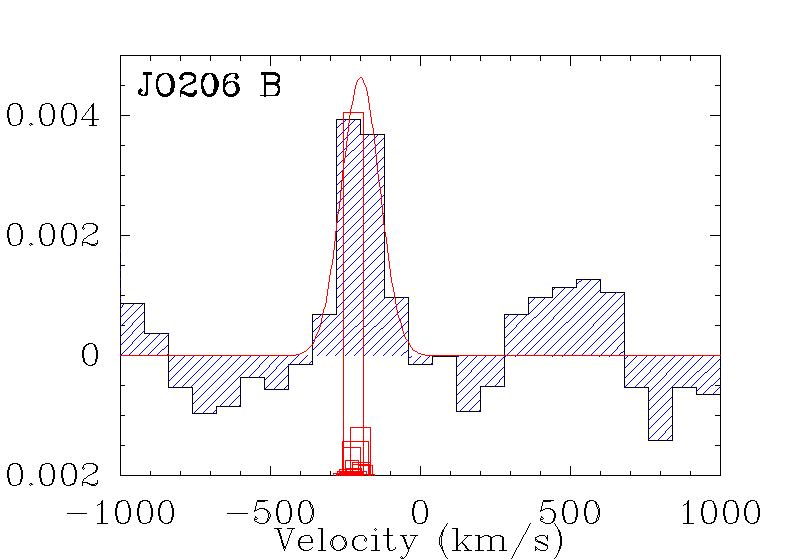}
\includegraphics[width=0.45\textwidth]{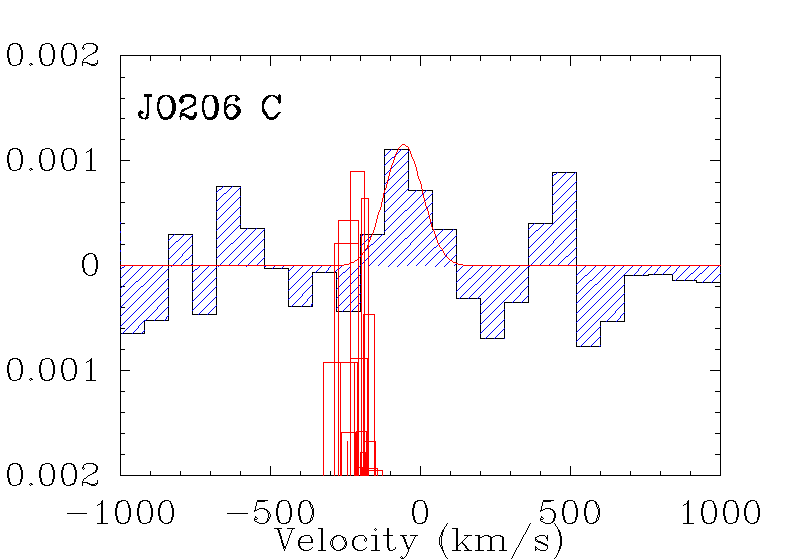}
\includegraphics[width=0.45\textwidth]{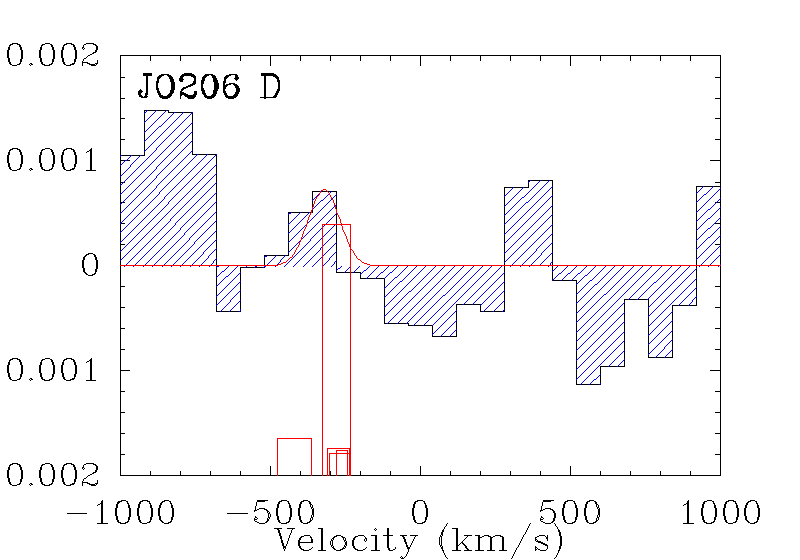}
\end{minipage}
\caption{{\bf JO206} The upper left panel shows the white light image of the galaxy extracted from the MUSE datacube. The upper right panel shows the stellar kinematics derived from MUSE spectra in Voronoi binned regions with $S/N=10$. The lower left panel shows the galaxy emission determined from the continuum under the $\rm H\alpha$ emission from the MUSE data (in grey) and the star forming knots color coded according to their velocity with respect to the galaxy center. Red (superimposed to the main galaxy body) and blue circles are the observed APEX pointings. The lower right panels show the CO($2-1$) spectra ($T^*_{mb}$) in the 4 positions. In each spectrum, the underlying red boxes corresponds to the $\rm H\alpha$ emissions of the star forming regions within the APEX beam (see text).\label{fig:JO206_1G}}
\end{figure*}

JO206 is the GASP galaxy showing the longest ionized gas tails \citep{gaspI} where 139 knots of star formation has been revealed through their $\rm H\alpha$ emission. With the exception of the central knot showing line ratios typical of AGN emission, almost all the other knots are consistent with being ionized by young stars formed in situ. Their median ionized gas mass is $\sim 1.5 \times 10^5 M_{\odot}$, and their total mass is $\sim 1.7 \times 10^8 M_{\odot}$, not including the diffuse $\rm H\alpha$ emission and the regions where the gas density could not be estimated using the [SII] 6716/[SII] 6732 ratio.
Our detections of the CO line, shown in Fig. \ref{fig:JO206_1G}, are significant ($S/N >3$) for the central pointing and for the region located at $\sim$ 30 kpc from the center (pointing B).
A free gaussian fit in the C and D pointings was not able to recover the molecular line, and we had to impose a peak position (-55 kms for the C pointing, -320 kms for the D pointing) in order to get the fit. If real, in C and D again the molecular gas appears to be at a different velocity with respect to the $\rm H\alpha$ emission.
In any case the C pointing is only marginally significant ($S/N=2.3$, $T_{ON}=93$ min), and the farthest D pointing (at $\sim 67$ kpc) is consistent with a non detection. We notice, however, that the observations of both these pointings are  shorter than what we asked for and longer exposures would be needed to ascertain the amount of molecular gas beyond $\sim$ 40 kpc from the galaxy main body.

\subsection{JW100}\label{sec:JW100}
\begin{figure*}
\centering
\begin{minipage}{0.45\textwidth}
\includegraphics[width=1.0\textwidth]{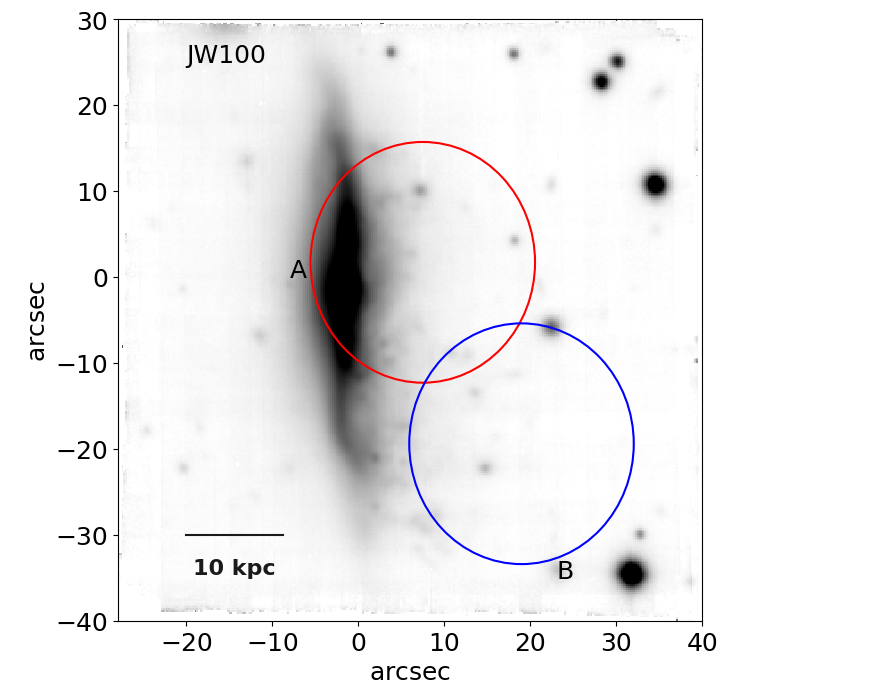}
\end{minipage}
\begin{minipage}{0.45\textwidth}
\includegraphics[width=1.0\textwidth]{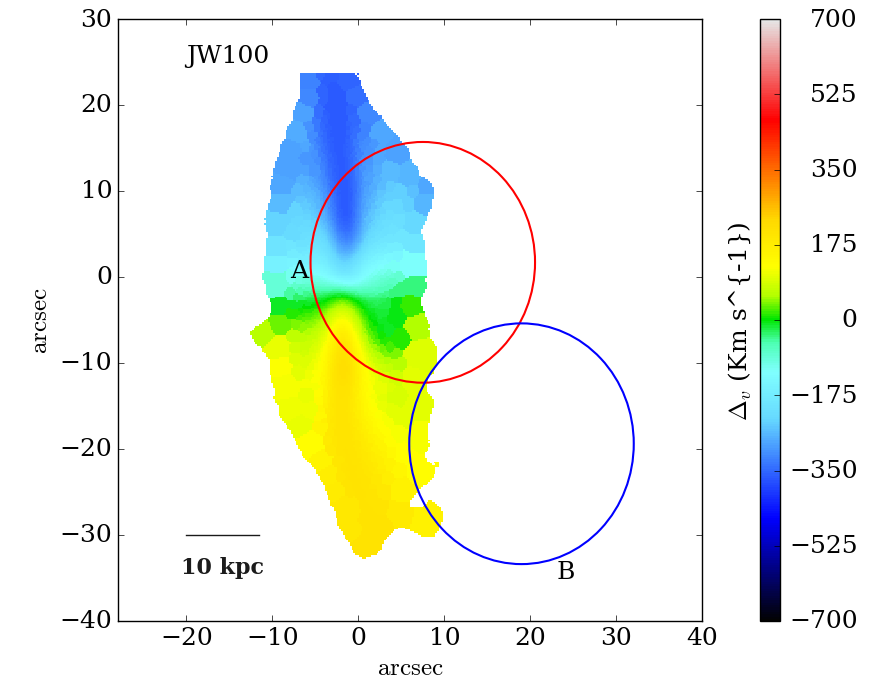}
\end{minipage}
\begin{minipage}{0.45\textwidth}
\includegraphics[width=1.0\textwidth]{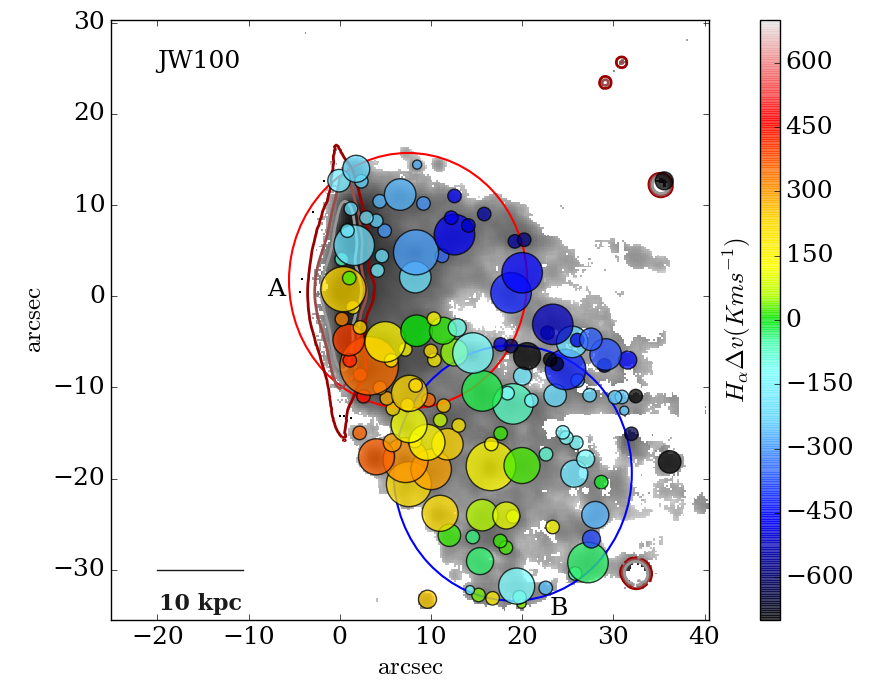}
\end{minipage}
\begin{minipage}{0.45\textwidth}
\includegraphics[width=0.45\textwidth]{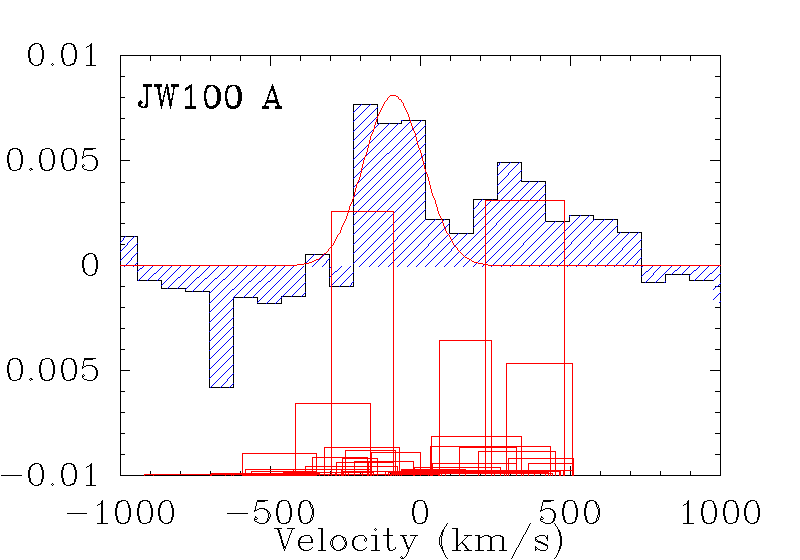}
\includegraphics[width=0.45\textwidth]{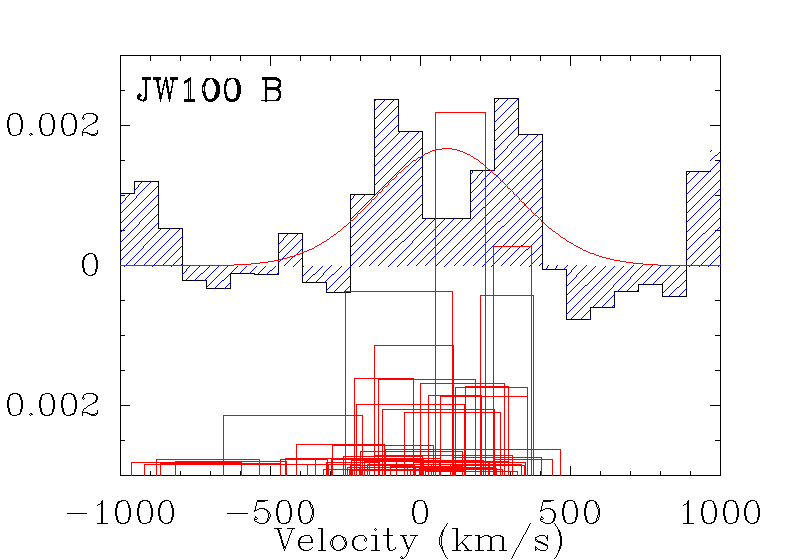}
\includegraphics[width=0.45\textwidth]{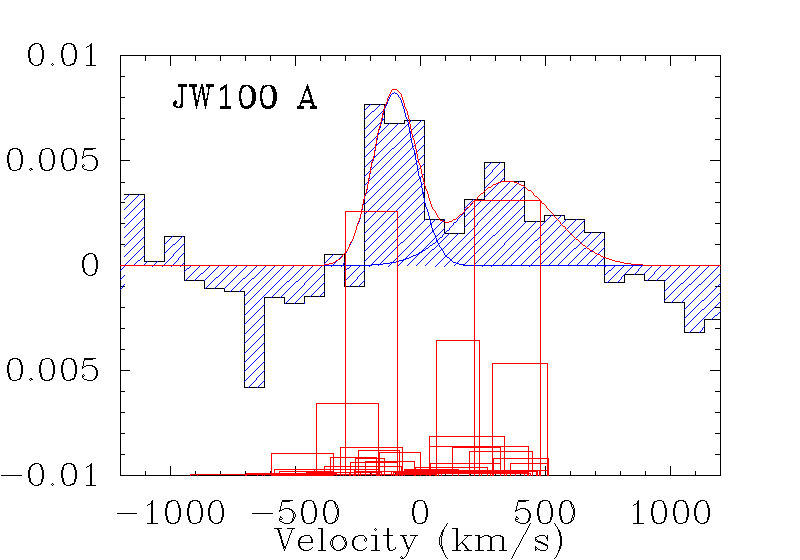}
\includegraphics[width=0.45\textwidth]{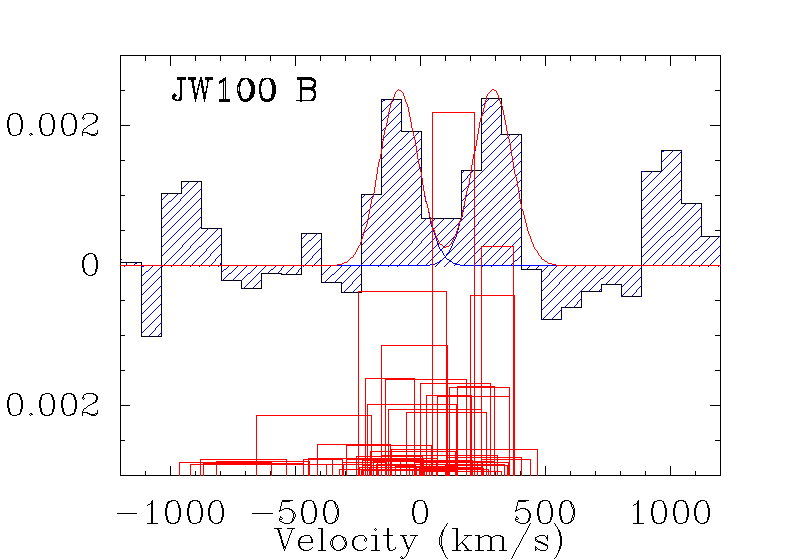}
\end{minipage}
\caption{{\bf JW100} The upper left panel shows the white light image of the galaxy extracted from the MUSE datacube. The upper right panel shows the stellar kinematics derived from MUSE spectra in Voronoi binned regions with $S/N=10$. The lower left panel shows the galaxy emission determined from the continuum under the $\rm H\alpha$ emission from the MUSE data (in grey) and the star forming knots color coded according to their velocity with respect to the galaxy center. Red (superimposed to the main galaxy body) and blue circles are the observed APEX pointings. The lower right panels show the CO($2-1$) spectra ($T^*_{mb}$) in the 2 positions (top row), while the bottom row show the 2-Gaussian fit of the emission line spectra. In each spectrum, the underlying red boxes corresponds to the $\rm H\alpha$ emissions of the star forming regions within the APEX beam (see text).
\label{fig:JW100_1G}}
\end{figure*}

JW100 (also known as IC5337) is a galaxy located close to the center of the cluster A2626.
\citet{Gitti+2013} detected it in radio continuum using VLA and suggested it is a head-tail radio galaxy, whose radio morphology is therefore linked to the ram-pressure effects. 
We have selected it as a candidate jellyfish on the basis of its appearance in the B-band images of the WINGS survey \citep{Poggianti+2016,Fasano+2006,Moretti2014}.
It turns out, in fact, that this galaxy is a real jellyfish galaxy (Sanchez et al., in preparation), and it also possesses an AGN at its center \citep{poggianti2017}.
We identified 106 star forming knots (among the 131 total knots) with a median mass of $0.8 \times 10^6 M_{\odot}$, accounting for a total mass of ionized gas of $5.63 \times 10^7$ solar masses.
We observed it with APEX using the two pointings A and B shown in the left panel of Fig.\ref{fig:JW100_1G}, one (A) containing almost all the galaxy disk, and the other (B) along the most prominent tail.
As can be seen from Fig.\ref{fig:JW100_1G}, right panels, both pointings reveal the presence of molecular gas, and neither of them can be well fitted with a single Gaussian component.
Both pointings require two gaussian profiles, almost coincident with the two peaks of $\rm H\alpha$ emission and having similar CO fluxes (lower right panels in Fig.\ref{fig:JW100_1G}).
In this case the velocities of the two components of warm ionized gas and cold CO emitting gas are well correlated.

To summarize, in the tails of our jellyfish galaxies, the velocities of significant CO detections overlap (fully or in part) with the velocities of the $\rm H\alpha$ emitting knots within that beam (JO201 D and F, JO204 B and C, JO206 B, JW100 B.), except for the galaxy in our sample with a very strong line-of-sight velocity component, JO201, whose tentative CO detections (S/N$\sim 3$) in pointings B, C and E have no overlap in velocity between CO and $ \rm H\alpha$.

The generally good coincidence between CO and $\rm H\alpha$ emission is consistent with the 
molecular gas we detect in the tails representing the gas reservoir out of which star formation takes place. 
Additional CO velocity components in these pointings are most probably indicating a differential effect of the ram pressure stripping on the cold and warm gas.
Finally, the fact that the CO emission in the B, C and E JO201 pointings retains the stellar velocity at the closest position in the disk but has no associated $\rm H\alpha$ is puzzling. It might suggest this molecular gas has formed from gas that has been more recently stripped and has either not yet formed stars or will never form stars.

It is hard to assess based only on kinematical considerations whether the molecular gas in the tails was formed in situ from the condensation of stripped neutral gas or if it was stripped in molecular form. Given the expected lifetimes of molecular clouds (a few Myr, \citep{elme2000}) and the large galactocentric distances at which we observe the CO the most plausible hypothesis is in-situ formation within the stripped tails. 
Moreover, the average molecular gas content in spiral galaxies is $\sim 10\%$,\citep{Saintonge2011}, and this is consistent with our findings in the pointings encompassing the galaxy light (see Sec.{\ref{sec:h2masses}}), which means that no molecular gas mass is missing within these regions (see also \citet{Kenney+1989} for similar results in the Virgo cluster), making it very plausible that the molecular gas in the tails has been formed in-situ.

\section{$\rm H_2$ masses}\label{sec:h2masses}
We derived the $H_2$ masses using the formulation by \citet{WatsonKoda2016}, i.e.
\begin{equation}
\left(\frac{M_{H_2}}{M_{\odot}}\right) = 
3.8 \times 10^3 \left( \frac{\alpha_{10}}{4.3}\right)
\left(\frac{r_{21}}{0.7}\right)^{-1}
\left(\int S_{21}dv \right)
\left(D \right)^2
\end{equation}
where 
$\alpha_{10}$ is the CO-to-$H_2$ conversion factor expressed in $M_{\odot}pc^{-2}$ \citep{Bolatto2013}, $r_{21}$ is the CO $J=2-1/1-0$ line ratio, $S_{21}$ is the CO integrated line flux in Jy and D is the distance in Mpc.

In calculating molecular gas masses we used $r_{21}=0.79$, as derived from the xCOLD GASS survey \citep{Saintonge2017} that studies a sample of 532 galaxies in the low-z Universe with masses larger than $10^9M_{\odot}$. 
The value adopted is slightly larger than the usual value of 0.7 derived from resolved observations of nearby star forming disks \citep{Leroy2013}, but lower than the value found in galaxies nuclei ($r_{21}\sim1$, \citealt{Leroy2009}).
As for the CO-to-$H_2$ conversion factor, we used both the standard Milky Way value of 4.3 (already including the helium content, \citealt{Leroy2011}) (9th column of Tab.\ref{tab:res}) and the one derived assuming the analytical dependence on the metallicity found in \citet{Accurso2017}, neglecting the small dependence on the distance from the Main Sequence of Star Formation (i. e. the locus of the star forming galaxies in the SFR versus $M_{\star}$ plane), since our values do not refer to the whole galaxy (10th column of Tab.\ref{tab:res}).
This relation is valid in the range $7.9<(12+log(O/H))<8.8$, while most of the gas in our APEX pointings turned out to possess a metallicity larger than 8.8. We therefore used the value derived assuming 12+log(O/H)=8.8 (where $\alpha_{CO}=2.95$), as suggested by \citet{Accurso2017}.
In fact our galaxies have all large metallicities/masses and we expect them not to suffer from the usual problems related to the conversion factor in low metallicities galaxies.

In all galaxies we were able to measure significant amounts of cold gas both in the galaxy main body and along the tails: the central pointings reveal an $H_2$ mass that goes from $8.29 \times 10^9 M_{\odot}$ in JO204 to $19.17 \times 10^9 M_{\odot}$ in JW100.
Along the tails we find an upper limit of $0.63 \times 10^9 M_{\odot}$ of $H_2$ in the farthest position of JO206, while in general a few $10^9 M_{\odot}$ of molecular gas are found within each APEX beam.

Looking at the ratio between the total measured mass of molecular hydrogen and the total galaxy stellar mass  we find that it is always smaller than one (
0.42, 0.20, 0.14 in JO204, JO206 and JW100, respectively 
) except in JO201, where we detect a mass of molecular gas similar to the stellar mass of the galaxy. However, as described in Sec. \ref{sec:results}, a significant fraction of this molecular gas could not be associated with the stripped tails, meaning that a not negligible fraction of it is still linked to the galaxy disks.

Our measured ratios are in agreement with studies of normal spiral galaxies \citep{Leroy2008,Saintonge2011} and with the predictions from semi-analytical models \citep{Popping+2014}.

\begin{table*}
\centering
 \label{tab:res}
\begin{tabular}{lllllllllll}
Gx & pointing & RMS & S/N & Rel. vel. &$T_{mb}$ & FWHM & CO flux  & $M(H_2)$MW&$M(H_2)$\\
   &          & mK  &     & Km/s      &mK       & km/s  & Jy Km/s  & $10^9M_{\odot}$&$10^9M_{\odot}$\\
\hline
JO201 & A    & 0.9 & 5.1 & 46   &5.0 &360.7 & 75.15 & 9.60&6.59\\
\hline
JO201 & B    & 0.5 & 3.0 & -207 &1.7 &363.0 & 25.55 & 3.26&2.24\\
\hline
JO201 & C    & 1.2 & 2.1 & -107 &2.6 &510.4 & 54.44 &6.96&5.69\\
	  & C1   &     & 2.4 & -177 &3.2 &278.7 & 36.74 &4.69&3.84\\
	  & C2   &     & 1.3 & 154  &1.7 &200.0* & 14.48 &1.85&1.51\\
\hline
JO201 & D    & 0.9 & 2.9 & 390  &2.7 &548.7 & 61.74 &7.89&5.41\\
      & D1   &     & 2.2 & -6   &2.2 &249.9 & 22.56 &2.88&1.98\\
      & D2   &     & 3.4 & 484  &3.4 &336.7 & 47.34 &6.05&4.15\\
\hline
JO201 & E    & 0.9 & 3.6 & 12   &3.3 &139.4 & 19.18 &2.45&1.72\\
      & E1   &     & 3.4 & 11   &3.4 &128.5 & 18.35 &2.35&1.64\\
      & E2   &     & 0.8 & 300*  &0.8 &221.9 & 7.42  &0.95&0.66\\
\hline
JO201 & F    & 0.6 & 4.0 & 344  &2.5 &400.7 & 41.57 &5.31&3.64\\
      & F1   &     & 4.1 & 298 & 2.4 &344.6 & 34.66 &4.43&3.04\\
      & F2   &     & 3.6 & 529 & 2.1* &80.6  & 7.03  &0.90&0.62\\
\hline
JO204 & A    & 1.0 & 4.1 & -26  &4.4 &392.8 & 72.11 &8.29&5.69\\
\hline
JO204 & B    & 0.8 & 3.1 & 67   &3.3 &333.9 & 45.79 &5.26&3.61\\
      & B1   &     & 3.1 & -26  &2.6 &229.3 & 25.13 &2.89&1.98\\
      & B2   &     & 3.2 & 160* &2.7 &171.2 & 19.30 &2.22&1.52\\
\hline
JO204 & C    & 0.6 & 3.2 & -45  &2.0 &353.8 & 28.87 &3.32&2.28\\
      & C1   &     & 3.2 &-120* &2.0 &167.3 & 14.24 &1.64&1.12\\
      & C2   &     & 2.2 &83    &1.4 &232.2 & 13.36 &1.54&1.05\\
\hline
JO206 & A    & 1.0 & 4.3 & 45   &4.9 &302.8 & 61.46 &10.43&7.16\\
\hline
JO206 & B    & 0.9 & 4.7 & -198 &4.6 &151.2 & 29.14 &4.95&3.39\\
\hline
JO206 & C    & 0.5 & 2.3 & -55*   &1.2 &149.7 & 7.19 &1.22&1.13\\
\hline
JO206 & D    & 0.7 & 0.9 & -320*   &0.7 &123.7 & 3.73 &0.63&0.91\\
\hline
JW100 & A    & 1.8 & 3.5 & -89  & 8.1 & 233.4 & 78.97 &19.17&13.15\\
      & A1   & 2?  & 3.6 & -102 & 8.2 & 204.1 & 69.86 &16.96&11.63\\
      & A2   &     & 1.8 & 350  & 4.0 & 427.2 & 71.71 &17.41&11.94\\
\hline
JW100 & B    & 0.9 & 2.2 & 89   & 1.7 & 540.4 & 37.60 &9.13&6.26\\
      & B1   & 0.7 & 3.2 & -84  & 2.5 & 180.5 & 18.86 &4.58&3.14\\
      & B2   &     & 3.2 & 291  & 2.5 & 184.8 & 19.33 &4.69&3.22\\
\hline
\end{tabular}
\caption{For each APEX pointing we give the 1 $\sigma$ RMS in mK, the signal to noise (S/N), the peak CO velocity in km s$^{-1}$, the main beam temperature in mK, the FWHM in km s$^{-1}$ and the total CO flux within the fitted line in Jy km s$^{-1}$. The last two columns refer to the corresponding $H_2$ mass calculated assuming the Milky Way CO-to-$H_2$ conversion factor and the metallicity dependent CO-to-$H_2$ conversion factor from \citet{Accurso2017}.}
\end{table*}

\section{Star Formation Efficiency} \label{sec:sfe}
The most widely used way to identify the conditions at which gas is able to form stars is to evaluate the so called Star Formation Efficiency (SFE), which is the ratio between the Star Formation Rate (SFR) surface density and the $H_2$ surface density.
This SFE is the inverse of the $H_2$ gas depletion time, i.e. the time needed to consume the available $H_2$ at the current SFR.
This means, for example, that if 1\% of the $H_2$ is converted into stars every $10^8$ years, then the SFE is $10^{-10}$ yr$^{-1}$.

In order to estimate the SFE in the different APEX pointings we used the molecular gas masses derived in Sec. \ref{sec:h2masses}  normalized to the APEX beam area (HPBW), and the resolved SFR inferred from the MUSE datacubes as follows.
We first corrected the MUSE datacubes for the stellar absorption modeled using the SINOPSIS code \citep{gaspIII}, and then run the KUBEVIZ code \citep{Fossati2016} on the emission-only MUSE datacube 
to measure all the relevant emission lines that have been used to i) estimate the Balmer decrement to correct for the dust extinction, and ii) classify each spaxel on the basis of the ionization mechanism using the BTP diagram \citep{BPT}, following the procedure described in \citet{gaspI}.

\begin{figure}
\centering
\includegraphics[width=0.45\textwidth]{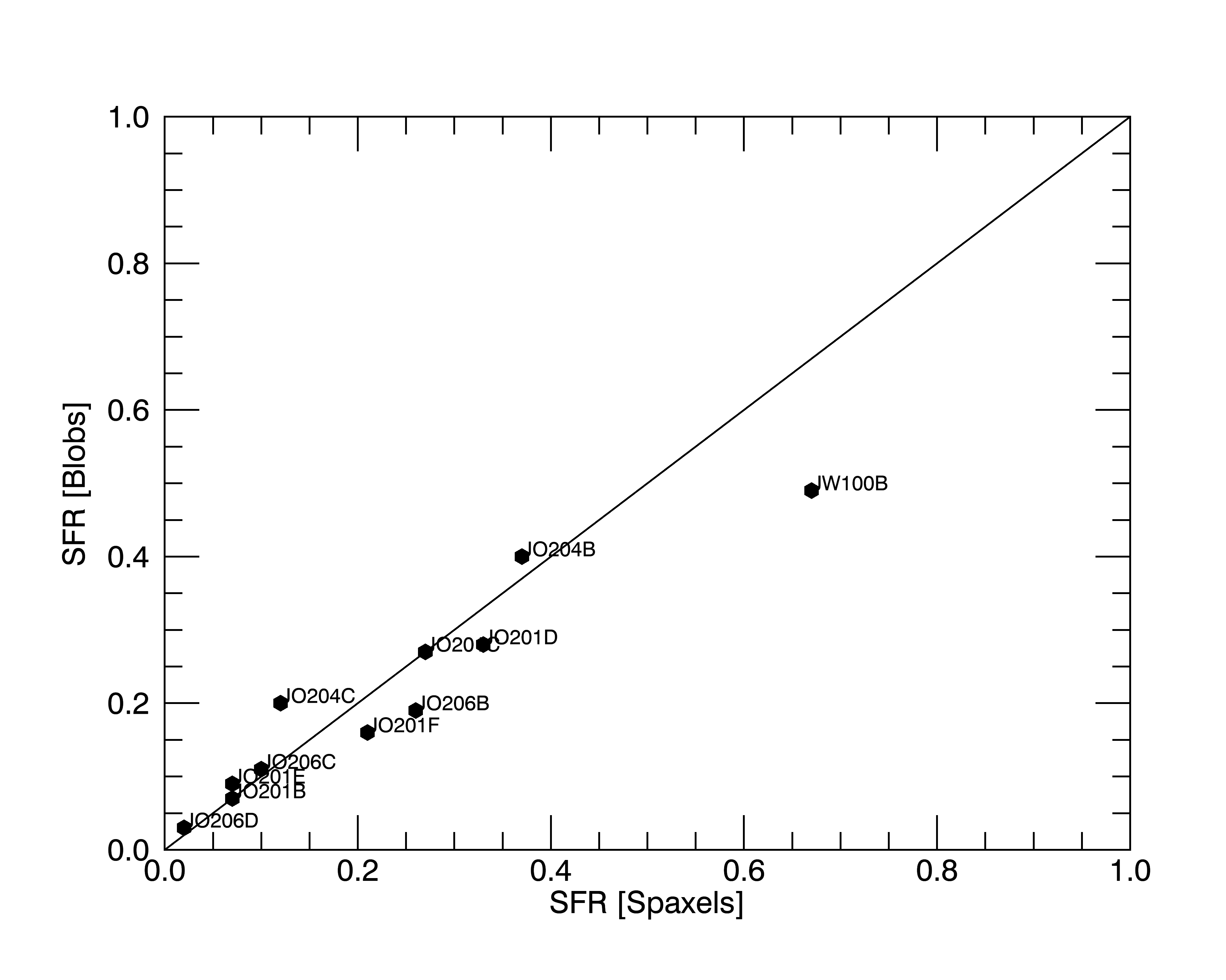}
\caption{Comparison between the Star Formation Rate measured from the star forming regions and from the single star forming spaxels within each APEX pointing\label{fig:cfr_sfr}}
\end{figure}

\begin{figure}
\centering
\includegraphics[width=0.45\textwidth]{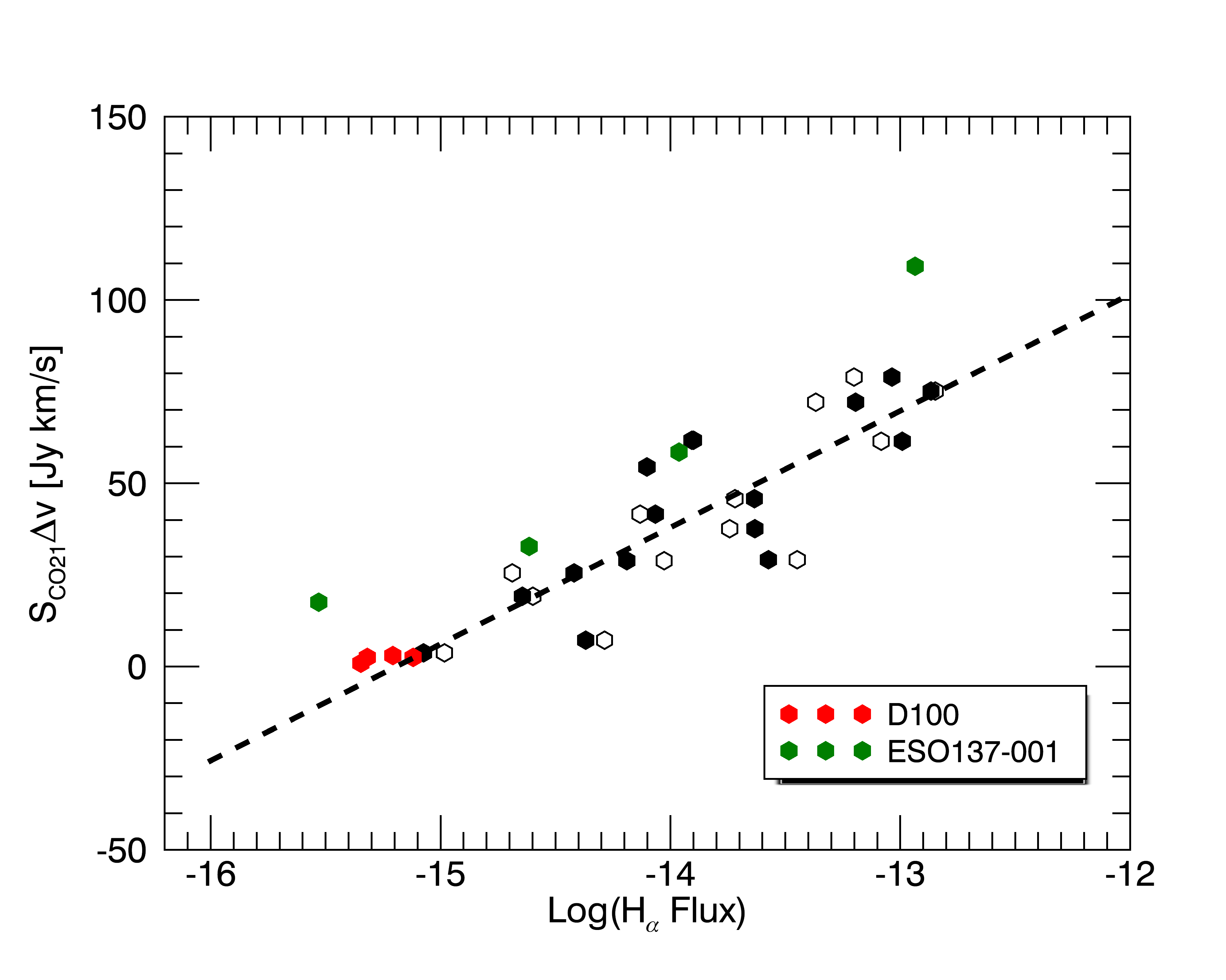}
\caption{CO fluxes from APEX data against $\rm H\alpha$ fluxes (in $erg s^{-1}$) derived from GASP/MUSE observations: black filled symbols refer to $\rm H\alpha$ fluxes measured from individual spaxels, empty symbols to $\rm H\alpha$ fluxes measured from the sum of individual star forming knots, in both cases classified as star forming or composite. Red and green symbols are literature values for other two nearby jellyfish galaxies, D100 and ESO137-001, respectively, from \citet{Jachym2017} and \citet{Jachym2014}. \label{fig:sco_hasb}}
\end{figure}

The SFR within each APEX pointing has been computed from the $\rm H\alpha$ luminosity (corrected for dust and stellar contribution) using the Kennicutt (\citeyear{Kennicutt1998})'s relation
\begin{equation}
SFR = 4.6 \times 10^{-42}L_{\rm H\alpha}
\label{eqn:sfr}
\end{equation}
where the SFR (for a \citealt{Chabrier2003} IMF) is given in solar masses per year and the $\rm H\alpha$ luminosity is in erg per second.
In these calculations we only used spaxels classified as either pure star forming or located in the so called Composite region in the $O[III]/H_{\beta}$ versus $N[II]/\rm H\alpha$ BPT diagram, using the demarcation lines by \citet{Kewley2001} and \citet{Kauffmann2003}.
We also estimated the total SFR within each APEX beam by summing the contribution of each star forming knot found using the procedure described in \citet{gaspI}, finding very similar values (see Fig. \ref{fig:cfr_sfr}). This means that most of the star formation in the galaxy tails takes place within the knots.
Before evaluating the SFE, which depends on the assumptions on the $\alpha_{CO}$ and the $r_{21}/r_{10}$ factors, we compared the $\rm H\alpha$ fluxes measured from the MUSE data within the beam with the total CO flux that we measured.
This is shown in Fig.\ref{fig:sco_hasb}, where black dots refer to our data, while red dots are values taken from \citep{Jachym2017} for different pointings in the D100 galaxy in the Coma cluster and green dots are relative to the ESO137-001 galaxy from \citet{Jachym2014}.
The empty symbols indicate the $\rm H\alpha$ flux that we measure from the HII regions identified with the MUSE data.
We find that a linear relation links  the CO fluxes (in $Jy\cdot km s^{-1}$) in our stripped tails and the $\rm H\alpha$ luminosities (in $erg s^{-1}$), i.e.
\begin{equation}
S_{CO21}\Delta v = 483.3+31.8\times Log(\rm H\alpha)
\end{equation}
This is true both when we consider the integrated $\rm H\alpha$ luminosity of each spaxel classified as HII region or composite, and if we consider the contribution of each individual HII region within the APEX beam.
This relation holds independently of the galaxy masses, given the large range of galaxy masses involved in the plot (from the $2 \times 10^9$ of D100 to $3 \times 10^{11}$ in JW100).
This means that the mechanism producing the ionizing flux (probably Star Formation) is strictly linked to the molecular gas available.

In contrast, the galaxy stellar mass seems to be comparable to the total molecular gas mass only for JO201, hosted in a cluster with $M> 10^{15} M_{\odot}$. 
The other galaxies, instead, reside in clusters with smaller masses
($M_{200}\sim 0.4 \times 10^{15} M_{\odot}$, \citet{Moretti+2017}) and their molecular gas is smaller ($1.7 \times 10^{10} M_{\odot}$ for JO204 and JO206 and $4.4\times 10^{10} M_{\odot}$ for JW100) than the stellar
mass.
We have been able to roughly estimate the efficiency of the RPS for three of our galaxies (JO201, JO204 and JO206) by making standard assumptions on the stellar/gas densities in our galaxies and assuming a simple model for the clusters ICM \citep{gaspIX}. It turns out that galaxies where the stripping is more efficient (JO201 with $\sim$50\% of its gas being stripped) also possess the largest quantities of molecular gas mass, while the ones where the gas stripping has been less efficient (JO204 and JO201, with $\sim$40\% and $\sim$15\% of gas stripped, respectively) have less molecular hydrogen. 

\begin{figure}
\centering
\includegraphics[width=0.45\textwidth]{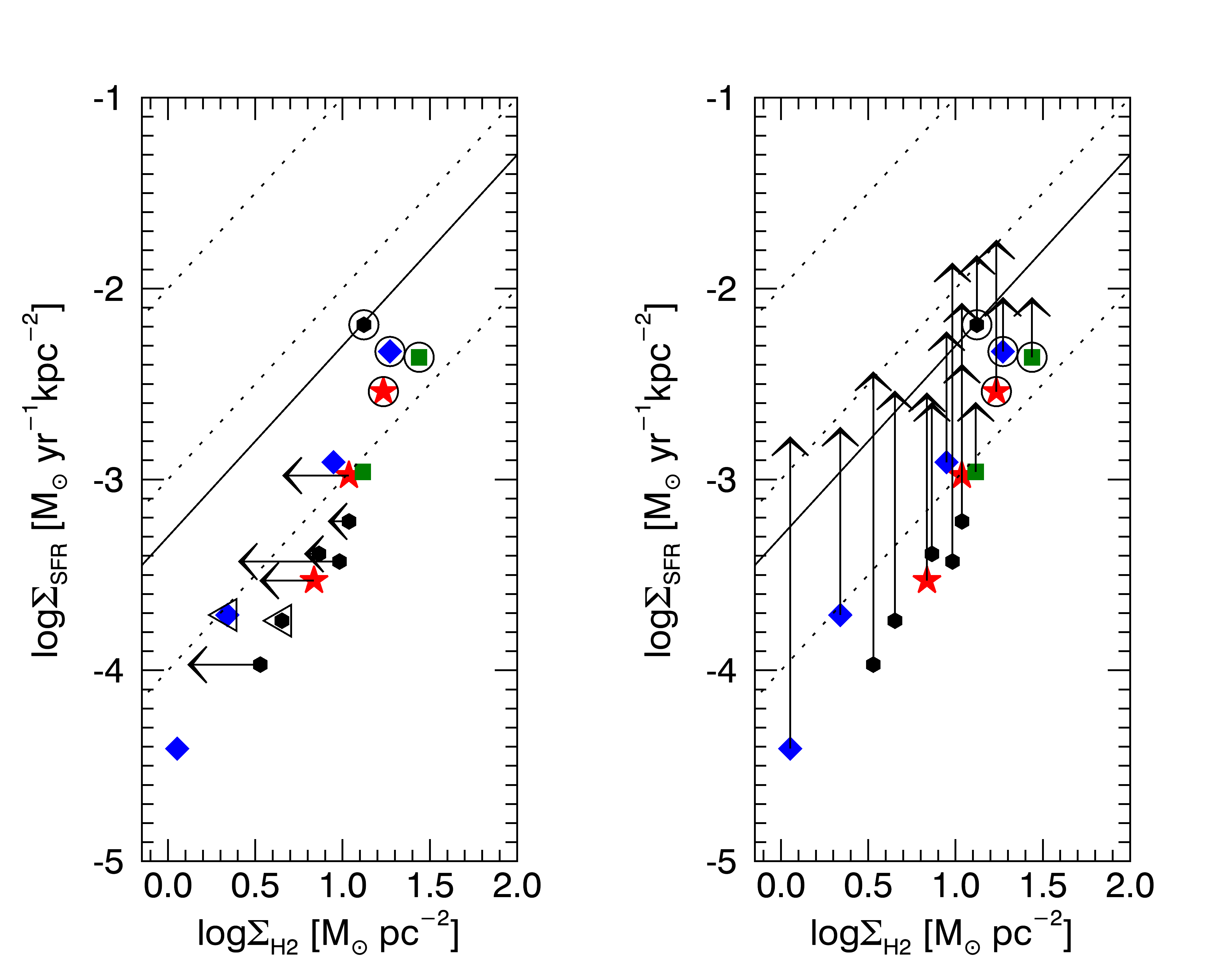}
\caption{Star formation rate surface density, $\Sigma_{SFR}$, derived following Eqn.\ref{eqn:sfr} as a function of molecular gas surface density, $\Sigma_{H_2}$, estimated from CO(2-1) emission from APEX for the different pointings along the tails and in the main body of the jellyfish galaxies: black dots refer to JO201, red stars to JO204, blue diamonds to JO206 and green squares to JW100. The four symbols showing the highest SFRs indicate the measurements in the main body of our galaxies (also enclosed within a circle). Dotted lines represent fixed $H_2$ depletion times in years ($10^8$, $10^9$ and $10^{10}$ from top to bottom), while the continuous line shows the average depletion time found for a sample of 30 nearby disk galaxies with a resolution of 1 kpc  \citep{Bigiel2011}. Arrows in the left panel show molecular gas surface densities considering only the molecular gas coincident with the $\rm H\alpha$ emission. The two triangles indicate pointings whose the molecular gas surface density associated with the $\rm H\alpha$ emission extend beyond the plot limits. Arrows in the right panel show SFRs densities calculated using the effective area covered by the $\rm H\alpha$ emission.\label{fig:SFE}}
\end{figure}

Having converted the CO fluxes into molecular gas masses we calculated the surface densities of $H_2$ within each APEX beam and compare them with the SFR surface density, that is shown in both panels of Fig.\ref{fig:SFE} illustrating the Kennicutt-Schmidt relation \citep{Schmidt1959,Kennicutt1998}.
Different symbols and colors have been used for different galaxies: black dots are for JO201, red stars for JO204, blue diamonds for JO206 and green squares for JW100. The symbols enclosed within circles indicate the central pointings of each galaxy.
The dotted lines in Fig.\ref{fig:SFE} represent constant depletion times of $10^8$, $10^9$, and $10^{10}$ years from top to bottom, while the continuous line shows the average depletion time ($\sim 2.3$ Gyr) found by \citet{Bigiel2011} for a sample of 30 nearby disk galaxies (IRAM HERACLES CO survey).
The SFR densities that we measure in the stripped tails are very low, or equivalently the gas depletion times are very long, meaning that it will take longer than the Hubble time to consume the molecular gas in the tails with the current SFR.
Our results are in agreement with what has been found in other ram-pressure stripped galaxies \citep{Jachym2014,Verdugo2015,Jachym2017}, and imply that most of the molecular gas in the tails will not be used to fuel the star formation, but will ultimately get dispersed into the ICM.

However, while the APEX beam size does not allow to resolve the location of the CO emitting region, we do have from MUSE data the ability to distinguish the $\rm H\alpha$ location, with a resolution of $\sim1$ kpc \citep{gaspI}. Therefore, we can make two working hypothesis: either (a) the $\rm H\alpha$ and CO emission come from co-spatial gas phases, and in this case we should calculate a partially resolved CO emission using only the CO component that is coincident in velocity with the $\rm H\alpha$ peaks, or (b) the two phases might not be strictly linked (due to a more diffuse CO or $\rm H\alpha$ emission), and in this case we can evaluate each indicator with respect to the covered area that we can estimate with our data. This means assuming a diffuse CO emission over the whole APEX beam, and a clumpy $\rm H\alpha$ emission traced by the star forming knots.
The arrows in the left panel of Fig. \ref{fig:SFE} illustrate case (a), and indicate the $H_2$ surface densities obtained using the masses corresponding to the gaussian fit of the CO line coincident in velocity with the $\rm H\alpha$ knots. The two triangles indicate pointings where the $H_2$ surface density calculated in this way extends beyond the plot limits.
The arrows in the right panel of Fig.\ref{fig:SFE} refer, instead, to the case (b) i.e. using the effective area covered by the $\rm H\alpha$ emission in evaluating the SFR surface density (while the $H_2$ surface density does not change).
In both cases the extremely low SFE characterizing the stripped tails are brought back to the depletion times expected for normal star forming galaxies.
In other words, our results seem to indicate that the star formation can proceed normally within the stripped tails, but only a resolved study on the CO emission will definitively answer this question.
We will discuss this possibility in a forthcoming paper studying the same sample of jellyfish galaxies with ALMA observations.

\begin{figure}
\centering
\includegraphics[width=0.45\textwidth]{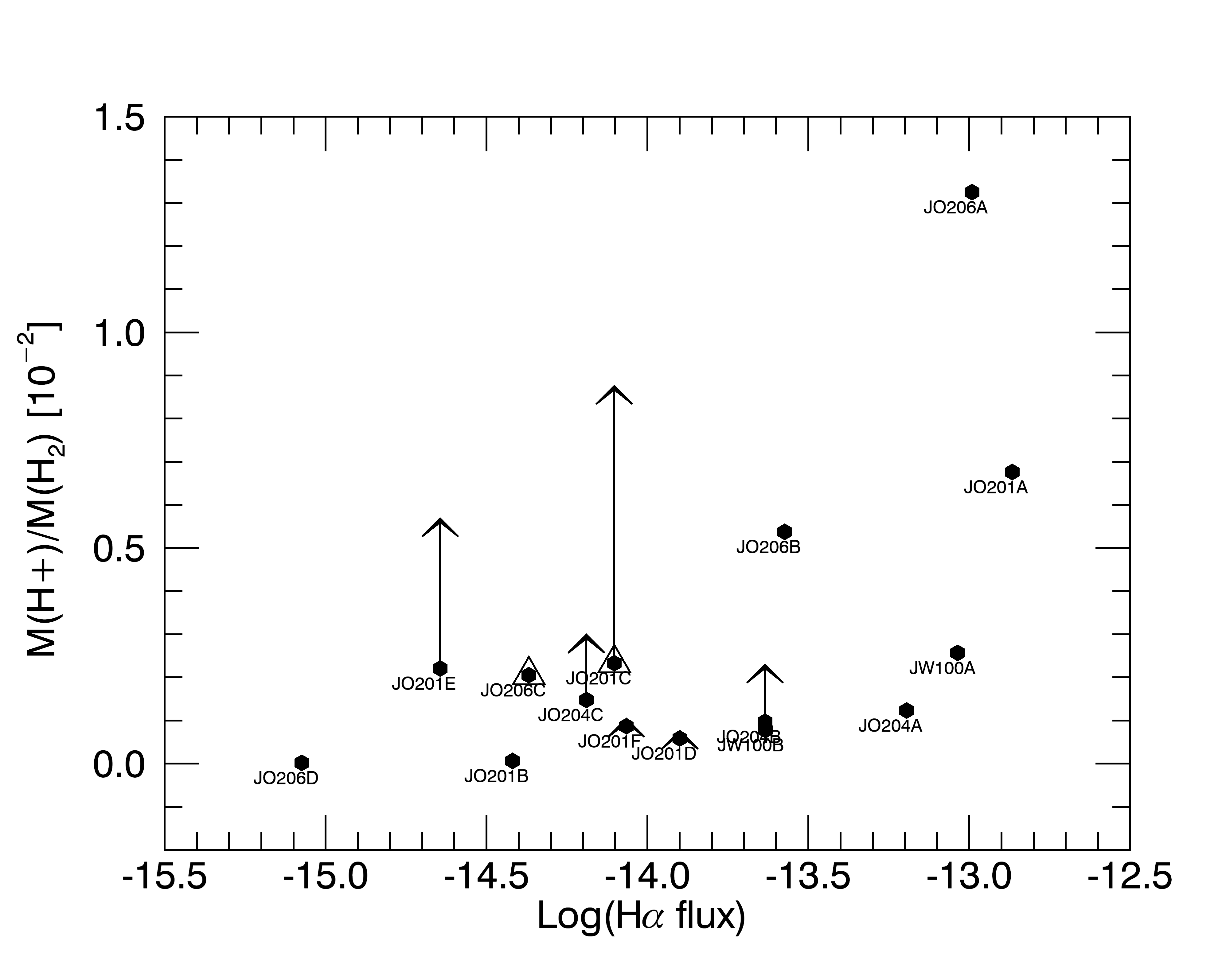}
\caption{Ratio between the mass of ionized and molecular hydrogen within each APEX beam versus the total $\rm H\alpha$ flux within the beam. The arrows indicate the ratios obtained by using only the $H_2$ masses associated with the $\rm H\alpha$ emission. Symbols within the triangles indicate pointings whose ratio between the two gas masses extends beyond the plot limits.\label{fig:mion_h2}}
\end{figure}

Finally we can calculate the total amount of ionized gas using the $\rm H\alpha$ fluxes measured with MUSE.
In order to get the total mass of ionized gas within the APEX beam we used Eqn. 3 from \citet{gaspI},
 that requires an estimate of the electron density {\it $n$}.
To derive it, we used the ratio between the red SII lines at 6716\, \AA \, and 6732\, \AA \,(excluding spaxels with a signal to noise below 3 in both lines) and the calibration by \citet{Proxauf+2014} (valid in the range $0.4<[SII]6716/[SII]6732<1.435$). 
Moreover, since SF is known to take place within HII regions, we did not derive the electron density for each spaxel, but for each HII region/complex identified within the APEX beam.
We then sum up all the evaluable masses in each beam to obtain the total mass of ionized gas to be compared with the $H_2$ mass. Fig.\ref{fig:mion_h2} shows that the ratio between the mass already ionized by star formation and the available molecular gas mass within each APEX beam is $\sim0.1-0.2 \times 10^{-2}$, which is one order of magnitude smaller than the one found by \citet{Jachym2017} along the tail of a smaller jellyfish galaxy (even though the central regions of JO201 and JO206 show more compatible results).
Our determinations, however, are based only on star forming regions where the SII doublet was measurable and within the range of the adopted calibration, therefore our values have to be taken as lower limits.
Moreover, if we consider only the $H_2$ mass associated with the $\rm H\alpha$ emission, the results are shifted towards higher values, as indicated by the arrows.

\section{Summary and conclusions}
We have studied a sample of 4 jellyfish galaxies from the GASP sample \citep{gaspI} with the APEX telescope to derive the molecular gas content both in the galaxy main bodies and along the stripped tails through the CO(2-1) emission.
Similar studies have been performed so far only in three nearby jellyfish galaxies \citep{Jachym2014,Verdugo2015,Jachym2017} in Norma, Virgo, and Coma clusters, respectively.
Our galaxies lie at higher redshifts ($\sim 0.05$), and are characterized by larger stellar masses.
All of them are galaxies subject to peak stripping \citep{gaspIX}, even though with a different stripping efficiency.
Our APEX CO observations demonstrates that: 
\begin{itemize}
\item large amounts of molecular gas are present in the disk of our jellyfish galaxies ($8-20 \times 10^9 M_{\odot}$), compatible with the values expected in normal star forming galaxies \citep{Saintonge2011}
\item also the pointings located along the stripped tails show a significant amount of molecular gas up to the largest distances covered by our observations (from 20 to 67 kpc)
\item the total mass of molecular gas that we measure for each galaxy by summing all the pointings goes from 15\% to 100\% of the galaxy stellar mass measured from the integrated spectrum within the stellar isophote encircling the galaxy main body as traced by the continuum under the $\rm H\alpha$ emission down to 1$\sigma$ above the background of our MUSE data cubes.
\item there is a clear correlation between the CO(2-1) fluxes and the $\rm H\alpha$ fluxes measured within each beam indicating that the ionized emission and the amount of molecular gas are strictly linked
\item there is a shift between the $\rm H\alpha$ and the CO velocities in the tails, probably due to the differential effect of the ram pressure on cold and warm gas. This conclusion is however only preliminary, since the APEX beam contains many $\rm H\alpha$ knots, each one with its own velocity, therefore hampering a clear correspondence between the two gas phases. 
\item the central regions of each galaxy (encircled points in Fig.\ref{fig:SFE}) show average SFEs, corresponding to the typical depletion times found for nearby disks \citep{Leroy2011}, while along the stripped tails the situation dramatically changes: the SFE is very low and the depletion times imply that most of the molecular gas will not be converted into stars before joining the ICM, confirming the results already found in other nearby jellyfish galaxies \citep{Jachym2014,Verdugo2015,Jachym2017}.
However, our SFR derivation is based on the resolved $\rm H\alpha$ emission (on a kpc scale) from the MUSE data, and it is therefore possible to normalize the SFR to the effective area covered by this emission. This would then represent the limit case in which the CO emission is diffuse within the entire APEX beam, while only part of it is contributing to the star formation that is concentrated in the smaller HII regions/complexes traced by $\rm H\alpha$ emission \citep{Shetty+2014,Mogotsi+2016}.
Moreover, the joint kinematic analysis of the CO and the resolved $\rm H\alpha$ emission suggests that our values of SFE are consistent with the ones shown by normal disk galaxies, if we consider only the molecular gas associated with the warm phase.
This will be probed directly using our ongoing resolved ALMA observations of the CO emission.
\end{itemize}

This work demonstrates that galaxies subject to the ram pressure stripping effect in nearby clusters of galaxies (at redshift $\sim$ 0.05) show tails of multicomponent gas, where the molecular gas is not negligible and act as the needed reservoir to form a new generation of stars that is linked to the ionized gas emission.
Given that the central pointings reveal the presence of molecular gas masses compatible with the one of normal disk galaxies, we suggest that the molecular gas in the tails should have been formed {\it in situ} from the stripped neutral gas.
ALMA observations of these galaxies (undergoing) will help clarifying whether this scenario is confirmed.

\section*{Acknowledgements}

We thank Carlos De Breuck, Palle M\o\,ller and the APEX team for their support.
Based on observations collected by the European Organisation for Astronomical Research in the Southern Hemisphere under ESO program 098.B-0657 and 099.B-0063 (APEX) and 196.B-0578 (VLT/MUSE). 
We acknowledge funding from the INAF PRIN-SKA 2017 program 1.05.01.88.04.
B.V. acknowledges the support from an Australian Research Council Discovery Early Career Researcher Award (PD0028506).
This work made use of the KUBEVIZ software, which is publicly available at \url{http://www.mpe.mpg.de/~dwilman/kubeviz/}.




\bibliographystyle{mnras}
\bibliography{Mendeley} 



\bsp	
\label{lastpage}
\end{document}